\documentclass[preprint,authoryear,12pt]{elsarticle}
\usepackage[left=2.6cm,right=1.7cm, height = 22cm]{geometry}
\usepackage{amssymb,amsmath,amstext,psfrag}
\usepackage{accents,  hyperref}

\journal{Some Elsevier Journal}

\begin{document}

\begin{frontmatter}

\title{Ductile damage model for metal forming simulations including refined description of void nucleation}

\address[ad1]{Lavrentyev Institute of Hydrodynamics, Department of Solid Mechanics, Pr. Laverentyeva 15, 630090 Novosibirsk, Russia}
\address[ad2]{Novosibirsk State University, Pirogova 2, 630090 Novosibirsk, Russia}
\address[ad3]{Technische Universit\"at Chemnitz,
Department of Solid Mechanics, Reichenhainer Strasse 90, 09126 Chemnitz, Germany}

\author[ad1,ad2]{A. V. Shutov \corref{cor1}}
\cortext[cor1]{Corresponding author.
Tel.: +7 383 333 17 46; fax: +7 383 333 16 12.}
\makeatletter
\ead{alexey.v.shutov@gmail.com}
\makeatother
\author[ad3]{C. B. Silbermann}
\author[ad3]{J. Ihlemann}

\begin{abstract}

We address the prediction of
ductile damage and material anisotropy
accumulated during plastic deformation of metals. A new model of phenomenological
metal plasticity is proposed which is suitable for applications
involving large deformations of workpiece material.
The model takes combined nonlinear isotropic/kinematic hardening,
strain-driven damage and rate-dependence of the
stress response into account. Within this model,
the work hardening and the damage evolution are fully coupled.
The description of the kinematics is based on the
double multiplicative decomposition of the
deformation gradient proposed by Lion.
An additional multiplicative decomposition is introduced in order to
account for the  damage-induced volume increase of the material.
The model is formulated in a thermodynamically admissible manner.
Within a simple example of the proposed framework,
the material porosity is adopted as a rough measure of damage.

A new simple void nucleation rule is formulated based
on the consideration of various nucleation mechanisms.
In particular, this rule is
suitable for materials which exhibit a higher void nucleation
rate under torsion than in case of tension.

The material model is implemented into the FEM code Abaqus and a simulation of a
deep drawing process is presented. The robustness of the algorithm and the performance
of the formulation is demonstrated.

\end{abstract}


\end{frontmatter}



\newpage

\section*{Nomenclature}
\begin{tabbing}
$\mathbf{1}$   \quad \quad \quad \quad \quad \quad    \=  idenity tensor \\
$\mathbf{F}$, $\accentset{\text{por}}{\mathbf{F}}_{\text{ep}}$  \> deformation gradient and its elasto-plastic part (cf. \eqref{1stDecomp}) \\
$\accentset{\text{por}}{\mathbf F}_{\text{i}}$, $\accentset{\text{por}}{\mathbf F}_{\text{ii}}$ \>  dissipative parts of deformation (cf. \eqref{2ndDecomp}, \eqref{3rdDecomp}) \\
$\hat{\mathbf{F}}_{\text{e}}$, $\check{\mathbf{F}}_{\text{ie}}$  \>  conservative parts of deformation (cf. \eqref{2ndDecomp}, \eqref{3rdDecomp}) \\
$\mathbf{C}$, $\accentset{\text{por}}{\mathbf{C}}_{\text{ep}}$ \>  total and elasto-plastic right Cauchy-Green tensors (cf. $\eqref{RCGT1}_1$)\\
$\accentset{\text{por}}{\mathbf{C}}_{\text{i}}$,
$\accentset{\text{por}}{\mathbf{C}}_{\text{ii}}$, \>  (dissipative) tensors of right Cauchy-Green type (cf. $\eqref{RCGT1}_2$, $\eqref{RCGT122}_1$)\\
$\hat{\mathbf{C}}_{\text{e}}$, $\check{\mathbf{C}}_{\text{ie}} $ \>  (conservative) tensors of right Cauchy-Green type (cf. $\eqref{RCGT1}_3$, $\eqref{RCGT122}_2$) \\
$\mathbf{L}$ \>  velocity gradient tensor \\
$\hat{\mathbf{L}}_{\text{i}}$ \>  inelastic velocity gradient (cf. $\eqref{InelasticFlow}_1$)\\
$\check{\mathbf{L}}_{\text{ii}}$ \> inelastic velocity gradient of substructure   (cf. $\eqref{InelasticFlowSubstruct}_1$) \\
$\mathbf{D}$ \>  strain rate tensor (stretching tensor) \\
$\hat{\mathbf{D}}_{\text{i}}$ \>  inelastic strain rate (cf. $\eqref{InelasticFlow}_2$) \\
$\check{\mathbf{D}}_{\text{ii}}$ \>  inelastic strain rate of substructure (cf. $\eqref{InelasticFlowSubstruct}_2$) \\
$s$, $s_{\text{d}}$, $s_{\text{e}}$ \>  inelastic arc-length and its parts (cf. \eqref{Odqvist}) \\
$\mathbf{T}$, $\widetilde{\mathbf{T}}$                \>  Cauchy stress and 2nd Piola-Kirchhoff stress \\
$\mathbf{S}$, $\mathbf{S}_{\text{ep}}$   \> Kirchhoff stress and Kirchhoff-like stress (cf. \eqref{KiechhoffEP})  \\
$\hat{\mathbf S}_{\text{ep}}$, $\accentset{\text{por}}{\mathbf{T}}_{\text{ep}} $ \>
pull-backs of $\mathbf{S}_{\text{ep}}$ to $\hat{\mathcal{K}}$ and $\accentset{\text{por}}{\mathcal{K}}$ (cf. \eqref{KiechhoffEP2}, \eqref{KiechhoffEP3}) \\
$\hat{\mathbf{X}}$, $\accentset{\text{por}}{\mathbf{X}}$,  $\check{\mathbf{X}}$ \>
backstresses on $\hat{\mathcal{K}}$, $\accentset{\text{por}}{\mathcal{K}}$, $\check{\mathcal{K}}$ (cf. \eqref{BackStressIntermed}, \eqref{BackStressPorous}, $\eqref{hyperelastic}_2$) \\
$\hat{\mathbf{\Sigma}}$, $\check{\mathbf{\Xi}} $ \> effective stress on
$\hat{\mathcal{K}}$ and Mandel-like backstress on $\check{\mathcal{K}}$ (cf. \eqref{EffectiveStress}, \eqref{MandelLikeBackstress}) \\
$R$                \>  isotropic hardening (stress) (cf. $\eqref{hyperelastic}_3$, \eqref{yieldF}) \\
$f$                \>  yield function (rate-dependent overstress) (cf. \eqref{yieldF}) \\
$\mathbf{A}^{\text{D}}$ \>  deviatoric part of a second-rank tensor \\
$\overline{\mathbf{A}}$ \>  unimodular part of a second-rank tensor (cf. \eqref{unim}) \\
$\mathbf{A}^{\text{T}}$ \>  transposition of a second-rank tensor \\
$\text{sym}(\mathbf{A})$  \>  symmetric part of a second-rank tensor \\
$ \| \mathbf{A} \| $       \> Frobenius norm of a second-rank tensor \\
$ \mathbf{A} : \mathbf{B} $  \> scalar product of two second-rank tensors \\
$\rho_{\scriptscriptstyle \text{R}}$, $\rho_{\scriptscriptstyle \text{por}} $ \> mass densities in
$\tilde{\mathcal{K}}$ and $\accentset{\text{por}}{\mathcal{K}}$ \\
$ \Phi$ \>  damage-induced volume change (cf. \eqref{PhiProp}) \\
$N$  \> void number per unit volume in $\tilde{\mathcal{K}}$ \\

\end{tabbing}

\section{Introduction}

Dealing with metal forming applications, it is often necessary to
asses the mechanical properties of the resulting engineering components,
including the remaining bearing capacity, accumulated defects, and residual stresses.
Thus, a state-of-the-art model for metal forming simulations should account
for various nonlinear phenomena. If the residual stresses and the spring back are of particular interest,
such a model should include the combined isotropic/kinematic hardening.
For some metals, however, the influence of ductile damage induced by plastic
deformation should be taken into account as well.
In spite of widespread applications involving large plastic deformations accompanied by
kinematic hardening and damage, only few material models cover these effects
(cf. \cite{SimoJu1989, Menzel2005, LinBrocks2006, Grammenoudis, BammannSolanki, BroeckerMatzenmiller2014}).

In the current study we advocate the approach to
plasticity/viscoplasticity based on the multiplicative
decomposition of the deformation gradient. This approach is gaining
popularity due to its numerous advantages like
the absence of spurious shear oscillations, the absence of non-physical dissipation
in the elastic range and the weak invariance under the change of the
reference configuration \citep{ShutovIhlemann2013}.
The main purpose of the current publication is to
promote the phenomenological damage modeling within the multiplicative framework.
Toward that end, the model of finite strain viscoplasticity proposed by \cite{Shutov1} is extended
to account for ductile damage.
The original viscoplastic model
takes the isotropic hardening of Voce type and the kinematic hardening
of Armstrong-Frederick type into account.
The model is based on a double multiplicative split of the deformation gradient, considered by
\cite{Lion2000}.\footnote{
The seminal idea of the double split was already used by several
authors to capture the nonlinear kinematic hardening
\citep{Helm, Tsakmakis2004, Dettmer2004,
Hartmann2008, HenannAnand, Brepols2013, ZHU2013, Brepols2014}.
It was also implicitly adopted by \cite{Menzel2005, Johansson2005}.
A simple extension to distortional hardening was presented by \cite{ShutovPanhansKr}. }
Both the original model and its extension are thermodynamically consistent.
We aim at the simplest possible extension which takes the following effects into account:
\begin{itemize}
\item nonlinear kinematic and isotropic hardening;
\item inelastic volume change due to damage-induced porosity;
\item damage-induced deterioration of elastic and hardening properties.
\end{itemize}
In this vein, we assume that the elastic properties deteriorate with increasing damage and remain isotropic
at any stage of deformation. The assumption of elastic isotropy is needed to exclude the
plastic spin from the flow rule, thus reducing the flow rule to six dimensions.

The interaction between damage and strain
hardening is explicitly captured by the extended model. It is known that the mechanisms of the
isotropic and kinematic hardening are not identical (cf. \cite{Barlat2003}).
Therefore, the isotropic and kinematic contributions to the overall hardening
deteriorate differently with increasing damage.
Concerning the coupling in the opposite direction, the accumulated plastic anisotropy has
a clear impact on the rate of damage accumulation upon the strain path change.

Since the proposed model is a generalization of the existing
viscoplasticity model, some well-established numerical
procedures can be used. The model is formulated as an open framework, suitable for
further extension. Physically motivated relations describing void nucleation, growth, and
coalescence can be implemented in a straightforward way, thus enriching the constitutive formulation.
The restrictions imposed on these relations by the second law of
thermodynamics are obtained in an explicit form.

A new refined void nucleation rule is introduced in this paper
for porous ductile metals with second phases.
This evolution law contains dependencies on the eigenvalues of the (effective)
stress. The rule is based
on consideration of various nucleation mechanisms,
like debonding of second phase particles under tensile and
shear loading or crushing of the particles under
high hydrostatic stress. Therefore, the material parameters
possess a clear mechanical interpretation, which
simplifies the parameter identification on the basis of microstructural observations
or molecular dynamics simulations.

The material model is validated using the experimental
flow curves presented by \cite{Horstemeyer1998} for a
cast A356 aluminium alloy as well as some experimental data on void
nucleation published by \cite{Horstemeyer2000}.

We close this introduction with a few remarks regarding notation.
The entire presentation is coordinate free (cf. \cite{ShutovKreissig2008Koordinate}).
Therefore, upright subscripts in the notation do not indicate tensorial index but stand for
different deformation mechanisms such as ``e" for \emph{elastic} and ``i" for \emph{inelastic}.
Throughout this article, bold-faced symbols denote first- and second-
rank tensors in $\mathbb{R}^3$.
Symbol $\mathbf 1$ stands for the identity tensor.
The deviatoric part of a tensor is given by $\mathbf{A}^{\text{D}} : = \mathbf{A} - \frac{1}{3} \text{tr}(\mathbf{A})  \mathbf 1$.
A scalar product of two second-rank tensors is denoted by $\mathbf{A} : \mathbf{B} = \text{tr}(\mathbf{A} \ \mathbf{B}^{\text{T}})$.
This scalar product gives rise to the Frobenius norm $ \| \mathbf{A} \| := \sqrt{\mathbf{A} : \mathbf{A} }$.
For a certain material particle, the material time derivative is denoted
by dot: $\dot{\mathbf{A}} := \frac{d}{d t} \mathbf{A}$, with the particle held fixed during differentiation.

\section{Material model}

\subsection{Finite strain kinematics}

\begin{figure}\centering
\psfrag{A}[m][][1][0]{\footnotesize energy}
\psfrag{B}[m][][1][0]{\footnotesize storage}
\psfrag{C}[m][][1][0]{\footnotesize dissipation}
\psfrag{D}[m][][1][0]{substructure}
\psfrag{I}[m][][1][0]{$\tilde{\mathcal{K}}$}
\psfrag{Q}[m][][1][0]{$\mathbf{F}_{\theta}$}
\psfrag{Z}[m][][1][0]{$\mathcal{K}^{\theta}$}
\psfrag{J}[m][][1][0]{$\check{\mathcal{K}}$}
\psfrag{K}[m][][1][0]{$\hat{\mathcal{K}}$}
\psfrag{R}[m][][1][0]{$\mathbf{F}_{\text{M}}$}
\psfrag{S}[m][][1][0]{$\mathcal{K}$} \psfrag{L}[m][][1][0]{$\mathbf
F_{\text{ii}}$} \psfrag{M}[m][][1][0]{$\mathbf F_{\text{i}}$}
\psfrag{N}[m][][1][0]{$\check{\mathbf F}_{\text{ie}}$}
\psfrag{O}[m][][1][0]{$\hat{\mathbf F}_{\text{e}}$}
\psfrag{P}[m][][1][0]{$\mathbf F$}
\psfrag{E}[m][][1][0]{\footnotesize expansion due}
\psfrag{F}[m][][1][0]{\footnotesize to porosity}
\psfrag{G}[m][][1][0]{elasto-plastic}
\psfrag{H}[m][][1][0]{deformation}
\scalebox{0.95}{\includegraphics{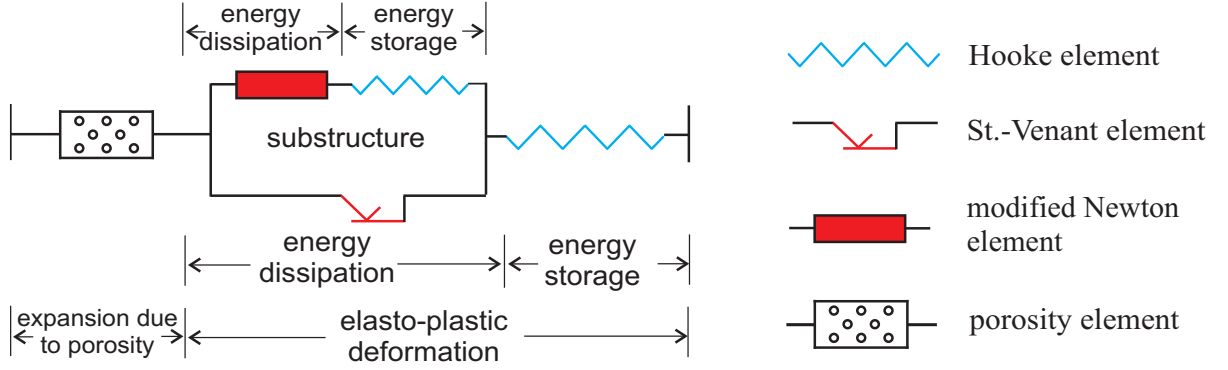}} \caption{Rheological
analogy: The model consists of an idealized porosity element connected in series with the modified Schwedoff element.
The modified Schwedoff element is build up of two elastic springs
(Hooke elements), one friction element (St.-Venant element), and a rate-independent
dashpot (modified Newton element).
\label{fig1}}
\end{figure}

In many cases, the consideration of rheological analogies provides insight into the main assumptions of material modeling
\citep{Reiner1960, Krawietz1986, Palmov1998, Palmov2008}.
As a starting point, we consider the rheological model shown in Fig. \ref{fig1}.
This model represents a modified Schwedoff element connected in series with an idealized porosity
element.\footnote{Note that this rheological model has a similar structure to that of
thermoplasticity with kinematic hardening
\citep{Lion2000, ShutovIhlemann2011}. In the current study, the element of thermal expansion
is replaced by the porosity element.}
The modified Schwedoff element exhibits nonlinear phenomena which are similar
to plasticity with Bauschinger effect \citep{Lion2000, Shutov1}.
The porosity element is introduced to account for
the effect of damage-induced volume change, which is important
for the correct prediction of hydrostatic component of stress.

\begin{figure}\centering
\psfrag{I}[m][][1][0]{$\tilde{\mathcal{K}}$}
\psfrag{Q}[m][][1][0]{$\mathbf{F}_{\text{por}}$}
\psfrag{Z}[m][][1][0]{$\accentset{\text{por}}{\mathcal{K}}$}
\psfrag{J}[m][][1][0]{$\check{\mathcal{K}}$}
\psfrag{K}[m][][1][0]{$\hat{\mathcal{K}}$}
\psfrag{R}[m][][1][0]{$\accentset{\text{por}}{\mathbf{F}}_{\text{ep}}$}
\psfrag{S}[m][][1][0]{$\mathcal{K}$}
\psfrag{L}[m][][1][0]{$\accentset{\text{por}}{\mathbf F}_{\text{ii}}$}
\psfrag{M}[m][][1][0]{$\accentset{\text{por}}{\mathbf F}_{\text{i}}$}
\psfrag{N}[m][][1][0]{$\check{\mathbf F}_{\text{ie}}$}
\psfrag{O}[m][][1][0]{$\hat{\mathbf F}_{\text{e}}$}
\psfrag{P}[m][][1][0]{$\mathbf F$}
\scalebox{0.95}{\includegraphics{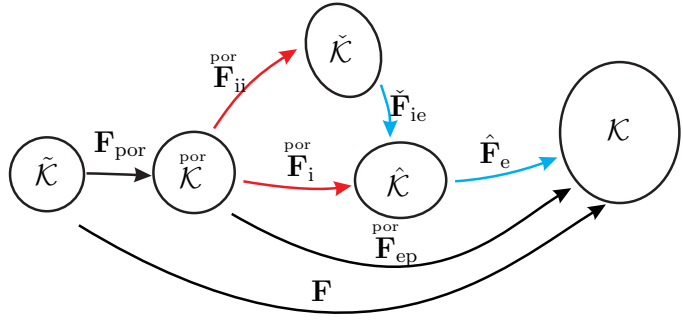}} \caption{Commutative diagram: triple multiplicative
decomposition of the deformation gradient.
\label{fig2}}
\end{figure}

In this paper we consider a system of constitutive equations
which qualitatively reproduces the rheological model.
Following the procedure described by \cite{Lion2000}, the displacements and forces in rheological elements
are formally replaced by strains and stresses. Any connection of rheological elements in series will be represented
by a multiplicative split of certain tensor-valued quantities.\footnote{This technique can
be successfully adopted even for two-dimensional rheological elements \citep{ShutovPanhansKr}.
For a general framework dealing with one-dimensional rheological
elements, we refer the reader to \cite{BroeckerMatzenmiller2014}.}
Let $\mathbf{F}$ be the deformation gradient from the local reference configuration
$\widetilde{\mathcal{K}}$ to the current configuration $\mathcal{K}$ (see Fig. \ref{fig2}).\footnote{A general
introduction to the kinematics within the geometrically
exact setting can be found in \cite{KhanHuang, Haupt, HashiguchiYamakawaBook}.}
First, we decompose it into an elasto-plastic part $\accentset{\text{por}}{\mathbf{F}}_{\text{ep}}$
and a porosity-induced expansion $\mathbf{F}_{\text{por}}$ in the following way
\begin{equation}\label{1stDecomp}
\mathbf{F} = \accentset{\text{por}}{\mathbf{F}}_{\text{ep}}  \mathbf{F}_{\text{por}},
\quad \mathbf{F}_{\text{por}} = \Phi^{1/3} \mathbf{1}, \quad \Phi \geq 1.
\end{equation}
Following \cite{BammannAifantis}, the expansion $\mathbf{F}_{\text{por}}$ is assumed to be isotropic,
where the damage-induced volume change is given by $ \Phi$
\begin{equation}\label{PhiProp}
\det \mathbf{F}_{\text{por}} = \Phi.
\end{equation}
The decomposition $\eqref{1stDecomp}_1$ yields a configuration of porous material
$\accentset{\text{por}}{\mathcal{K}} := \mathbf{F}_{\text{por}} \mathcal{K}$.
Let $\rho_{\scriptscriptstyle \text{R}}$ be the mass density in the reference configuration.
The mass density of unstressed porous material is given by $\rho_{\text{por}} := \rho_{\scriptscriptstyle \text{R}} / \Phi$.
Next, we consider the well-known decomposition of the elasto-plastic part into an
inelastic part $\accentset{\text{por}}{\mathbf F}_{\text{i}}$ and an elastic part $\hat{\mathbf{F}}_{\text{e}}$
\begin{equation}\label{2ndDecomp}
\accentset{\text{por}}{\mathbf{F}}_{\text{ep}} = \hat{\mathbf{F}}_{\text{e}}
\accentset{\text{por}}{\mathbf F}_{\text{i}}.
\end{equation}
\textbf{Remark 1.}
Originally, this decomposition was motivated for metallic materials by the idea of elastic unloading
to a stress-free state \citep{Bilby1957}.
In general, a pure elastic unloading is not possible and the multiplicative decomposition \eqref{2ndDecomp}
becomes a constitutive assumption.
There are some alternative ways to split the deformation into elastic and plastic
parts, including the additive split of the strain rate \citep{XiaoReview}.
It is worth mentioning that the multiplicative decomposition brings the advantage of the so-called weak invariance of
constitutive equations \citep{ShutovIhlemann2013}.  $\square$

Following the approach of \cite{Lion2000} to the nonlinear kinematic hardening, we decompose
the inelastic part into a dissipative part $\accentset{\text{por}}{\mathbf F}_{\text{ii}}$
and a conservative part $\check{\mathbf{F}}_{\text{ie}}$
\begin{equation}\label{3rdDecomp}
\accentset{\text{por}}{\mathbf F}_{\text{i}} = \check{\mathbf{F}}_{\text{ie}} \accentset{\text{por}}{\mathbf F}_{\text{ii}}.
\end{equation}
We adopt this decomposition as a phenomenological assumption, without explicit discussion of its microstructural origins \citep{Clayton2014}.
Intuitively, one can relate the dissipative parts $\accentset{\text{por}}{\mathbf F}_{\text{i}}$ and
$\accentset{\text{por}}{\mathbf F}_{\text{ii}}$ to the
deformations of the St.-Venant element and the modified Newton element, respectively (cf. Fig. \ref{fig1}).
The conservative parts $\hat{\mathbf{F}}_{\text{e}}$ and $\check{\mathbf{F}}_{\text{ie}}$
are associated with the deformation of the elastic springs (cf. Fig. \ref{fig1}).
The commutative diagram shown in Fig. \ref{fig2} is useful to summarize the introduced
multiplicative decompositions.
The fictitious local configurations $\hat{\mathcal{K}}$ and $\check{\mathcal{K}}$ are referred to
as stress-free intermediate configuration and configuration of kinematic hardening, respectively.

\textbf{Remark 2.} Combining $\eqref{1stDecomp}_1$ and \eqref{2ndDecomp} we arrive at
\begin{equation}\label{DecompTotal}
\mathbf{F} = \hat{\mathbf{F}}_{\text{e}} \accentset{\text{por}}{\mathbf F}_{\text{i}} \mathbf{F}_{\text{por}}.
\end{equation}
Due to the isotropy of $\mathbf{F}_{\text{por}}$, this decomposition is equivalent to
the decomposition considered by \cite{BammannAifantis, Ahad2014} and many others:
\begin{equation}\label{DecompTotal2}
\mathbf{F} = \hat{\mathbf{F}}_{\text{e}} \mathbf{F}_{\text{por}} \accentset{\text{por}}{\mathbf F}_{\text{i}}.
\end{equation}
In this paper we prefer to operate with \eqref{DecompTotal} rather than \eqref{DecompTotal2} in order to retain
the structure of the viscoplasticity model proposed by \cite{Shutov1}. $\square$

\textbf{Remark 3.}
Note that only a volumetric part of damage kinematics is explicitly captured by
$\mathbf{F}_{\text{por}}$ in this study.
Within a refined approach with enriched kinematic description, the damage-induced
isochoric deformation can be also considered explicitly.
It can be carried out through additional decomposition of
the inelastic part  $\accentset{\text{por}}{\mathbf F}_{\text{i}}$
into some isochoric damage-induced parts
and isochoric damage-free parts
(cf. \cite{VoyiadjisPark1999, Bruenig2003}).
Obviously, such an approach leads to an anisotropic damage model. $\square$

\textbf{Remark 4.}
There exists an abundance of literature considering
ductile damage modeling within multiplicative elasto-plasticity
based on the split \eqref{2ndDecomp} (see, for example, \\
\cite{BammannAifantis}; \cite{Steinmann1994}; \cite{Menzel2005}; \\
\cite{SoyarslanTekkaya2010};  \cite{BammannSolanki}, \cite{McAuliffe}).
Unfortunately, 
only few damage models incorporate
the multiplicative split \eqref{3rdDecomp} which is used
to capture the nonlinear kinematic hardening.
For instance, \cite{Grammenoudis, BroeckerMatzenmiller2014} have coupled ductile
damage with a model of finite strain plasticity by adopting
the concept of effective stresses combined with the strain equivalence principle.
Unfortunately, as will be seen from the following, such an approach is
too restrictive in some cases. In particular, it
implies that different hardening mechanisms deteriorate
with the same rate for progressive damage, which is in general not true (see Section 4).
Moreover, as already mentioned above, the damage-induced volume change has to be
explicitly taken into account for accurate prediction of the hydrostatic component of stress. $\square$

Let us consider the right Cauchy-Green tensor (RCGT) $\mathbf{C} := \mathbf{F}^{\text{T}} \mathbf{F}$.
Based on the decompositions $\eqref{1stDecomp}_1$ and \eqref{2ndDecomp}, we introduce
the following kinematic quantities of the right Cauchy-Green type
\begin{equation}\label{RCGT1}
\accentset{\text{por}}{\mathbf{C}}_{\text{ep}} := \accentset{\text{por}}{\mathbf{F}}^{\text{T}}_{\text{ep}}
\ \accentset{\text{por}}{\mathbf{F}}_{\text{ep}}, \
\accentset{\text{por}}{\mathbf{C}}_{\text{i}} := \accentset{\text{por}}{\mathbf{F}}^{\text{T}}_{\text{i}}
\ \accentset{\text{por}}{\mathbf{F}}_{\text{i}}, \
\hat{\mathbf{C}}_{\text{e}} := \hat{\mathbf{F}}^{\text{T}}_{\text{e}} \ \hat{\mathbf{F}}_{\text{e}}.
\end{equation}
Here, the elasto-plastic RCGT $\accentset{\text{por}}{\mathbf{C}}_{\text{ep}}$ and the inelastic RCGT $\accentset{\text{por}}{\mathbf{C}}_{\text{i}}$
operate on the porous configuration $\accentset{\text{por}}{\mathcal{K}}$; the elastic RCGT $\hat{\mathbf{C}}_{\text{e}}$
operates on the stress-free configuration $\hat{\mathcal{K}}$.
Next, on the basis of decomposition \eqref{3rdDecomp}, we define the inelastic RCGT of substructure $\accentset{\text{por}}{\mathbf{C}}_{\text{ii}}$ and
the elastic RCGT of substructure $\check{\mathbf{C}}_{\text{ie}}$
\begin{equation}\label{RCGT122}
\accentset{\text{por}}{\mathbf{C}}_{\text{ii}} := \accentset{\text{por}}{\mathbf{F}}^{\text{T}}_{\text{ii}} \ \accentset{\text{por}}{\mathbf{F}}_{\text{ii}}, \
\check{\mathbf{C}}_{\text{ie}} := \check{\mathbf{F}}^{\text{T}}_{\text{ie}} \ \check{\mathbf{F}}_{\text{ie}}.
\end{equation}
The tensors $\accentset{\text{por}}{\mathbf{C}}_{\text{ii}}$ and $\check{\mathbf{C}}_{\text{ie}}$ operate on
$\accentset{\text{por}}{\mathcal{K}}$ and $\check{\mathcal{K}}$ respectively. Combining \eqref{1stDecomp} and $\eqref{RCGT1}_1$, we have
\begin{equation}\label{RCGT2}
\accentset{\text{por}}{\mathbf{C}}_{\text{ep}} = \Phi^{-2/3} \mathbf{C}.
\end{equation}
In order to describe the inelastic flow we consider the inelastic velocity gradient $\hat{\mathbf{L}}_{\text{i}}$ and the inelastic strain rate
$\hat{\mathbf{D}}_{\text{i}}$, both operating on $\hat{\mathcal{K}}$
\begin{equation}\label{InelasticFlow}
\hat{\mathbf{L}}_{\text{i}} := \Big(\frac{d}{dt} \accentset{\text{por}}{\mathbf{F}}_{\text{i}}\Big) \
\accentset{\text{por}}{\mathbf{F}}^{-1}_{\text{i}}, \quad
\hat{\mathbf{D}}_{\text{i}} := \text{sym} \big( \hat{\mathbf{L}}_{\text{i}} \big).
\end{equation}
Analogously, the inelastic flow of the substructure is captured using the
following tensors operating on $\check{\mathcal{K}}$
\begin{equation}\label{InelasticFlowSubstruct}
\check{\mathbf{L}}_{\text{ii}} := \Big(\frac{d}{dt} \accentset{\text{por}}{\mathbf{F}}_{\text{ii}}\Big) \
\accentset{\text{por}}{\mathbf{F}}^{-1}_{\text{ii}}, \quad
\check{\mathbf{D}}_{\text{ii}} := \text{sym} \big( \check{\mathbf{L}}_{\text{ii}} \big).
\end{equation}
In terms of the rheological analogy shown in Fig. \ref{fig1}, the strain rates $\hat{\mathbf{D}}_{\text{i}}$
and $\check{\mathbf{D}}_{\text{ii}}$ can be associated with the
deformation rates of the St.-Venant element and the modified Newton element, respectively.

The accumulated  inelastic arc-length (Odqvist parameter) is a strain-like internal variable defined through
\begin{equation}\label{Odqvist}
s(t) := \int_{0}^{t} \dot{s}(\tau) d \tau ,
\quad \dot{s}:= \sqrt{\frac{\displaystyle 2}{\displaystyle 3}} \| \hat{\mathbf{D}}_{\text{i}} \|.
\end{equation}
Along with $s$ we introduce its dissipative part $s_{\text{d}}$ such
that $s_{\text{e}} : =s - s_{\text{d}}$ controls the isotropic hardening.\footnote{
For simplicity, the rheological model in Fig. \ref{fig1} does not include the isotropic hardening.
The reader interested in idealized rheological elements
of isotropic hardening is referred to \cite{BroekerMatz}.}

\subsection{Stresses}

Let $\mathbf{T}$ and $\mathbf{S} := \det (\mathbf{F}) \mathbf{T}$ be respectively the Cauchy and Kirchhoff stresses.
We introduce a Kirchhoff-like stress tensor operating on the current configuration $\mathcal{K}$
\begin{equation}\label{KiechhoffEP}
\mathbf{S}_{\text{ep}} := \det \big(\accentset{\text{por}}{\mathbf{F}}_{\text{ep}}\big) \mathbf{T} = \Phi^{-1} \mathbf{S}.
\end{equation}
The elastic pull-back of this tensor yields its counterpart operating on the stress-free configuration $\hat{\mathcal{K}}$
\begin{equation}\label{KiechhoffEP2}
\hat{\mathbf{S}}_{\text{ep}} := \hat{\mathbf{F}}^{-1}_{\text{e}} \mathbf{S}_{\text{ep}} \hat{\mathbf{F}}^{-\text{T}}_{\text{e}}.
\end{equation}
The 2nd Piola-Kirchhoff tensor is obtained applying a pull-back to $\mathbf{S}$
\begin{equation}\label{2ndPKdeinit}
\tilde{\mathbf{T}} := \mathbf{F}^{-1} \mathbf{S} \mathbf{F}^{-\text{T}}.
\end{equation}
Analogously, the elasto-plastic pull-back of $\mathbf{S}_{\text{ep}}$
yields a Kirchhoff-like tensor operating on the
porous configuration $\accentset{\text{por}}{\mathcal{K}}$. At the same time, this tensor
is identical to the inelastic pull-back of $\hat{\mathbf{S}}_{\text{ep}}$:
\begin{equation}\label{KiechhoffEP3}
\accentset{\text{por}}{\mathbf{T}}_{\text{ep}} := \accentset{\text{por}}{\mathbf{F}}^{-1}_{\text{ep}} \mathbf{S}_{\text{ep}}
\accentset{\text{por}}{\mathbf{F}}^{-\text{T}}_{\text{ep}} = \accentset{\text{por}}{\mathbf{F}}^{-1}_{\text{i}} \hat{\mathbf{S}}_{\text{ep}}
\accentset{\text{por}}{\mathbf{F}}^{-\text{T}}_{\text{i}} .
\end{equation}
The stress tensors $\tilde{\mathbf{T}}$ and $\accentset{\text{por}}{\mathbf{T}}_{\text{ep}}$ are related by
\begin{equation}\label{RelatBet2ndPK}
\tilde{\mathbf{T}} = \Phi^{1/3} \ \accentset{\text{por}}{\mathbf{T}}_{\text{ep}}.
\end{equation}
Next, let us consider the backstresses which are typically used to capture
the Bauschinger effect.\footnote{A backstresses-free approach
to the Bauschinger effect was presented by \cite{Barlat2011}.}
Intuitively, the backstresses
correspond to stresses in the modified Newton element which is a part
of the Schwedoff element shown in Fig. \ref{fig1}. Let $\check{\mathbf{X}}$ be
the backstress operating on $\check{\mathcal{K}}$, which is a stress measure
power-conjugate to the strain rate $\check{\mathbf{D}}_{\text{ii}}$. A backstress measure operating on
the stress-free configuration $\hat{\mathcal{K}}$ is obtained using a push-forward operation as follows
\begin{equation}\label{BackStressIntermed}
\hat{\mathbf{X}} := \check{\mathbf{F}}_{\text{ie}} \check{\mathbf{X}} \check{\mathbf{F}}^{\text{T}}_{\text{ie}}.
\end{equation}
Its counterpart operating on the porous configuration $\accentset{\text{por}}{\mathcal{K}}$ is given by
\begin{equation}\label{BackStressPorous}
\accentset{\text{por}}{\mathbf{X}} :=
\accentset{\text{por}}{\mathbf{F}}^{-1}_{\text{ii}} \check{\mathbf{X}} \accentset{\text{por}}{\mathbf{F}}^{-\text{T}}_{\text{ii}} =
\accentset{\text{por}}{\mathbf{F}}^{-1}_{\text{i}} \hat{\mathbf{X}} \accentset{\text{por}}{\mathbf{F}}^{-\text{T}}_{\text{i}}.
\end{equation}
A Mandel-like backstress tensor on $\check{\mathcal{K}}$ is defined through
\begin{equation}\label{MandelLikeBackstress}
\check{\mathbf{\Xi}} := \check{\mathbf{C}}_{\text{ie}} \check{\mathbf{X}}.
\end{equation}
For what follows, it is instructive to introduce the so-called effective stress operating on $\hat{\mathcal{K}}$
\begin{equation}\label{EffectiveStress}
\hat{\mathbf{\Sigma}} := \hat{\mathbf{C}}_{\text{e}} \hat{\mathbf{S}}_{\text{ep}} - \hat{\mathbf{X}}.
\end{equation}
In other words, the effective stress represents the difference between the Mandel-like stress
$\hat{\mathbf{C}}_{\text{e}} \hat{\mathbf{S}}_{\text{ep}}$ and the backstress $\hat{\mathbf{X}}$.
For this definition, a Kirchhoff-like stress $\mathbf{S}_{\text{ep}}$ is adopted instead of the Kirchhoff
stress $\mathbf{S}$ in order to keep the structure of the damage model close to the viscoplasticity
model presented by \cite{Shutov1}.
The effective stress represents the local force driving the inelastic deformation.
In terms of the rheological model, this stress corresponds
to the load acting on the St.-Venant element.\footnote{
A distinction should be made between the effective stress concept which is typically
used in continuum damage mechanics \citep{Rabotnov1969}
and the notion of effective stress adopted within plasticity with
kinematic hardening. In fact, the second notion is implemented in the
current study with respect to $\hat{\mathbf{\Sigma}}$.}

\subsection{Free energy}

According to \cite{Tvergaard1990}, the scalar porosity parameter is an adequate damage measure at
the initial stage of ductile damage process.
Only small pore volume fractions will be considered in this paper, such that
the applicability domain of the model is restricted to $\Phi -1 \ll 1$.
The material porosity will be seen as a rough measure of damage (cf. also \cite{BammannAifantis}).
Aiming at a thermodynamically consistent formulation, we postulate the free energy per unit mass in the form
\begin{equation}\label{freeen}
\psi=\psi(\hat{\mathbf C}_{\text{e}}, \check{\mathbf C}_{\text{ie}}, s_{\text{e}}, \Phi)=
\psi_{\text{el}} (\hat{\mathbf C}_{\text{e}}, \Phi) + \psi_{\text{kin}} (\check{\mathbf C}_{\text{ie}}, \Phi) +
\psi_{\text{iso}}(s_{\text{e}}, \Phi),
\end{equation}
where $\psi_{\text{el}}$ corresponds to the energy stored due to
macroscopic elastic deformations, components $\psi_{\text{kin}}$ and $\psi_{\text{iso}}$
correspond to the energy storage associated with kinematic and isotropic hardening.\footnote{
More precisely, $\psi_{\text{kin}} + \psi_{\text{iso}}$ is a part of the free energy stored in the defects
of the crystal structure. Remaining parts of the ``defect energy"  which are detached from
any hardening mechanism \citep{ShutovIhlemann2011} are neglected here.
An accurate description of the energy storage becomes especially
important within thermoplasticity with nonlinear kinematic
hardening \citep{Canadja2004, CanadjaMosler2011, ShutovIhlemann2011}.}
The dependence of $\psi_{\text{el}} $, $\psi_{\text{kin}}$, and $\psi_{\text{iso}}$ on $\Phi$
is introduced in order to capture the damage-induced deterioration of elastic and hardening properties, as
is common for metals.
In this study, the system of constitutive equations will be formulated in case of general
isotropic energy-storage functions $\psi_{\text{el}}(\hat{\mathbf C}_{\text{e}}, \Phi)$
and $\psi_{\text{kin}}(\check{\mathbf{C}}_{\text{ie}}, \Phi)$.
However, some concrete assumptions will be needed for the numerical computations. To be definite, we postulate
\begin{equation}\label{spec1}
\rho_{\scriptscriptstyle \text{R}}  \psi_{\text{el}}(\hat{\mathbf C}_{\text{e}}, \Phi)=
\frac{k(\Phi)}{2}\big(\text{ln}\sqrt{\text{det} \hat{\mathbf{C}}_{\text{e}}} \big)^2+
\frac{\mu(\Phi)}{2} \big( \text{tr} \overline{\hat{\mathbf{C}}_{\text{e}}} - 3 \big),
\end{equation}
\begin{equation}\label{spec2}
\rho_{\scriptscriptstyle \text{R}}  \psi_{\text{kin}}(\check{\mathbf{C}}_{\text{ie}}, \Phi)=
\frac{c(\Phi)}{4}\big( \text{tr} \overline{\check{\mathbf{C}}_{\text{ie}}} - 3 \big),
\quad \rho_{\scriptscriptstyle \text{R}} \psi_{\text{iso}}(s_{\text{e}}) =\frac{\gamma(\Phi)}{2} \ (s_{\text{e}})^2,
\end{equation}
where the overline $\overline{(\cdot)}$ denotes the unimodular
part of a second-rank tensor
\begin{equation}\label{unim}
\overline{\mathbf{A}}:=(\det \mathbf{A})^{-1/3} \mathbf{A}.
\end{equation}
Although the elastic properties of the damaged material
remain isotropic,
different deterioration rates have to be introduced for
the bulk modulus $k(\Phi)$ and the shear modulus $\mu(\Phi)$ \citep{Bristow1960}
\begin{equation}\label{DegradationElasticity}
k(\Phi) = k_0 \exp \big(-\text{BRR} \cdot (\Phi-1)\big), \quad
\mu(\Phi) = \mu_0 \exp \big(-\text{SRR} \cdot (\Phi-1)\big).
\end{equation}
Here, constant parameters BRR $\geq 0$ and SRR $\geq 0$ stand for  ``bulk reduction rate" and  ``shear reduction rate"
respectively.
The assumption \eqref{DegradationElasticity}
can reproduce the experimentally observed decrease of Poisson's ratio.
Analogously, for the degradation of the hardening mechanisms we postulate
\begin{equation}\label{DegradationHardening}
c(\Phi) = c_0 \exp \big(-\text{KRR} \cdot (\Phi-1)\big), \quad
\gamma(\Phi) = \gamma_0 \exp \big(-\text{IRR} \cdot (\Phi-1)\big),
\end{equation}
with KRR $\geq 0$ and IRR $\geq 0$.
Here, the parameter KRR stands for ``kinematic hardening reduction rate", and
IRR for ``isotropic hardening reduction rate".

As follows from \eqref{DegradationElasticity} and \eqref{DegradationHardening}, the material
parameters $k_0 > 0$, $\mu_0 >0$, $c_0 > 0$, and $\gamma_0 > 0$
correspond to the reference state with $\Phi=1$.

\textbf{Remark 5.} Even for $k = const$ and $\mu = const$, a weak damage-elasticity coupling is predicted by
\eqref{spec1} due to the mass density reduction. This type of weak coupling is also exhibited
by the well-known Rousselier model (cf., for example, \cite{Rousselier1987}). $\square$

Now we introduce the following relations of hyperelastic type
which express the stresses $\hat{\mathbf S}_{\text{ep}}$, the backstresses $\check{\mathbf X}$,
and the isotropic hardening $R$
as a function of introduced kinematic variables
\begin{equation}\label{hyperelastic}
\hat{\mathbf S}_{\text{ep}}= 2 \rho_{\scriptscriptstyle \text{por}}
\frac{\displaystyle \partial \psi_{\text{el}}(\hat{\mathbf{C}}_{\text{e}}, \Phi)}
{\displaystyle \partial \hat{\mathbf{C}}_{\text{e}}}, \
\check{\mathbf X}= 2 \rho_{\scriptscriptstyle \text{por}}
\frac{\displaystyle \partial
\psi_{\text{kin}}(\check{\mathbf{C}}_{\text{ie}}, \Phi)}
{\displaystyle \partial \check{\mathbf{C}}_{\text{ie}}}, \
R= \rho_{\scriptscriptstyle \text{por}} \frac{\displaystyle \partial \psi_{\text{iso}}
(s_{\text{e}}, \Phi)}{\displaystyle \partial s_{\text{e}}}.
\end{equation}
The relations $\eqref{hyperelastic}_1$ and $\eqref{hyperelastic}_2$ can be motivated by
the rheological model shown in Fig. \eqref{fig1}, if a hyperelastic response is assumed for
both elastic springs. On the other hand, these relations
will play a crucial role in the proof of the thermodynamic consistency (cf. Section \ref{CDI}).
Another implication of \eqref{hyperelastic} is that
$(\hat{\mathbf S}_{\text{ep}}, \hat{\mathbf{C}}_{\text{e}})$,
$(\check{\mathbf X}, \check{\mathbf{C}}_{\text{ie}})$, and
$(R, s_{\text{e}})$ form conjugate pairs. Originally, a hyperelastic
relation similar to $\eqref{hyperelastic}_1$ was introduced by \cite{SimoOrtiz1985}.
The advantage of such approach is that
a truly hyperelastic behavior is obtained in the elastic range and the principle of
objectivity is trivially satisfied.

\subsection{Clausius-Duhem inequality}\label{CDI}

Here we consider the Clausius-Duhem inequality, which states that the internal dissipation $\delta_{\text{i}}$ is always non-negative.
For isothermal processes studied in this paper, this inequality takes the reduced form
(cf., for instance, \cite{Haupt})
\begin{equation}\label{cld}
\delta_{\text{i}} := \frac{1}{\rho_{\scriptscriptstyle \text{R}}} \tilde{\mathbf T} : \dot{\mathbf E} - \dot{\psi} \geq 0.
\end{equation}
Let us specialize this inequality for the free energy given by \eqref{freeen}. First, we decompose the stress
power into a power of the hydrostatic stress component
on the damage-induced volume expansion and the
remaining part\footnote{A similar decomposition is obtained for some models of thermoplasticity due
to the similar description of kinematics \citep{Lion2000, ShutovIhlemann2011}.}
\begin{equation}\label{PowerDecomposition}
\frac{\displaystyle 1}{\displaystyle \rho_{\scriptscriptstyle \text{R}}} \tilde{\mathbf T} : \dot{\mathbf E} =
\frac{\displaystyle 1}{\displaystyle \rho_{\text{por}}}
\Big( \frac{\displaystyle \dot{\Phi}}{\displaystyle 3 \Phi }
\accentset{\text{por}}{\mathbf{T}}_{\text{ep}} : \accentset{\text{por}}{\mathbf{C}}_{\text{ep}}
+  \accentset{\text{por}}{\mathbf{T}}_{\text{ep}} : \frac{1}{2} \dot{\accentset{\text{por}}{\mathbf{C}}}_{\text{ep}}\Big).
\end{equation}
Recalling \eqref{RCGT2} and \eqref{RelatBet2ndPK} we obtain the following relation for the hydrostatic stress component
\begin{equation}\label{VolumePower}
\accentset{\text{por}}{\mathbf{T}}_{\text{ep}} : \accentset{\text{por}}{\mathbf{C}}_{\text{ep}} = \Phi^{-1} \ \tilde{\mathbf{T}} : \mathbf{C}.
\end{equation}
Let us take a closer look at the stress power
$\accentset{\text{por}}{\mathbf{T}}_{\text{ep}} : \frac{1}{2} \dot{\accentset{\text{por}}{\mathbf{C}}}_{\text{ep}}$.
As is usually done in multiplicative inelasticity (cf. equation (2.28) in \cite{Lion2000}
or equation (13.258) in \cite{Haupt}),
this quantity is split into two components. More precisely,
it follows from \eqref{2ndDecomp} that
\begin{equation}\label{TwoCompo}
\accentset{\text{por}}{\mathbf{T}}_{\text{ep}} : \frac{1}{2} \dot{\accentset{\text{por}}{\mathbf{C}}}_{\text{ep}} =
\hat{\mathbf S}_{\text{ep}} : \frac{1}{2} \dot{\hat{\mathbf{C}}}_{\text{e}} +
\big( \hat{\mathbf{C}}_{\text{e}} \hat{\mathbf S}_{\text{ep}} \big) : \hat{\mathbf{L}}_{\text{i}}.
\end{equation}
Since $\psi_{\text{el}}$ is isotropic, relation $\eqref{hyperelastic}_1$
implies that the tensors $\hat{\mathbf{C}}_{\text{e}}$ and $\hat{\mathbf S}_{\text{ep}}$
commute. In that case, the Mandel tensor $\hat{\mathbf{C}}_{\text{e}}
\hat{\mathbf S}_{\text{ep}}$ is symmetric and \eqref{TwoCompo} yields
\begin{equation}\label{TwoCompo2}
\accentset{\text{por}}{\mathbf{T}}_{\text{ep}} : \frac{1}{2} \dot{\accentset{\text{por}}{\mathbf{C}}}_{\text{ep}} =
\hat{\mathbf S}_{\text{ep}} : \frac{1}{2} \dot{\hat{\mathbf{C}}}_{\text{e}} +
\big( \hat{\mathbf{C}}_{\text{e}} \hat{\mathbf S}_{\text{ep}} \big) : \hat{\mathbf{D}}_{\text{i}}.
\end{equation}
Next, it follows from
$\eqref{hyperelastic}_2$ that $\check{\mathbf{C}}_{\text{ie}}$ and $\check{\mathbf X}$ commute
as a result of the isotropy assumption made for $\psi_{\text{kin}}$. Therefore, the
Mandel-like tensor $\check{\mathbf{\Xi}} = \check{\mathbf{C}}_{\text{ie}} \check{\mathbf{X}}$ is symmetric.
Taking the multiplicative split \eqref{3rdDecomp} into account, we obtain a relation for the backstress power,
which has a similar structure to \eqref{TwoCompo2}
\begin{equation}\label{TwoCompo3}
\hat{\mathbf{X}} : \hat{\mathbf{D}}_{\text{i}}  =
\check{\mathbf{X}} : \frac{1}{2} \dot{\check{\mathbf{C}}}_{\text{ie}} +
\big( \check{\mathbf{C}}_{\text{ie}} \check{\mathbf X} \big) : \check{\mathbf{D}}_{\text{ii}}.
\end{equation}
Rearranging the terms, we arrive at
\begin{equation}\label{TwoCompo322}
0 = - \hat{\mathbf{X}} : \hat{\mathbf{D}}_{\text{i}}  +
\check{\mathbf{X}} : \frac{1}{2} \dot{\check{\mathbf{C}}}_{\text{ie}} +
\big( \check{\mathbf{C}}_{\text{ie}} \check{\mathbf X} \big) : \check{\mathbf{D}}_{\text{ii}}.
\end{equation}
Summing both sides of \eqref{TwoCompo2} and \eqref{TwoCompo322} we obtain the stress power as
\begin{equation}\label{EPStressPower}
\accentset{\text{por}}{\mathbf{T}}_{\text{ep}} : \frac{1}{2} \dot{\accentset{\text{por}}{\mathbf{C}}}_{\text{ep}} =
\hat{\mathbf S}_{\text{ep}} : \frac{1}{2} \dot{\hat{\mathbf{C}}}_{\text{e}} +
\big( \hat{\mathbf{C}}_{\text{e}} \hat{\mathbf S}_{\text{ep}} \big) : \hat{\mathbf{D}}_{\text{i}} -
\hat{\mathbf{X}} : \hat{\mathbf{D}}_{\text{i}}  +
\check{\mathbf{X}} : \frac{1}{2} \dot{\check{\mathbf{C}}}_{\text{ie}} +
\big( \check{\mathbf{C}}_{\text{ie}} \check{\mathbf X} \big) : \check{\mathbf{D}}_{\text{ii}}.
\end{equation}
On the other hand, for the rate of the free energy we have
\begin{equation}\label{EnergRate}
\dot{\psi} \stackrel{\eqref{freeen}}{=}
\frac{\displaystyle \partial \psi_{\text{el}}(\hat{\mathbf{C}}_{\text{e}}, \Phi)}
{\displaystyle \partial \hat{\mathbf{C}}_{\text{e}}}: \dot{\hat{\mathbf{C}}}_{\text{e}} +
\frac{\displaystyle \partial \psi_{\text{kin}}(\check{\mathbf{C}}_{\text{ie}}, \Phi)}
{\displaystyle \partial \check{\mathbf{C}}_{\text{ie}}} : \dot{\check{\mathbf{C}}}_{\text{ie}}
 +  \frac{\displaystyle \partial \psi_{\text{iso}}
(s_{\text{e}}, \Phi)}{\displaystyle \partial s_{\text{e}}}  \dot{s}_{\text{e}} +
\frac{\displaystyle \partial \psi(\hat{\mathbf C}_{\text{e}},
\check{\mathbf C}_{\text{ie}}, s_{\text{e}}, \Phi)}{\displaystyle \partial \Phi} \dot{\Phi}.
\end{equation}
Substituting \eqref{PowerDecomposition} and \eqref{EnergRate} into \eqref{cld} and taking
\eqref{EPStressPower} into account we rewrite the internal dissipation in the form
\begin{multline}\label{dissi}
\delta_{\text{i}} =  \Big( \frac{\displaystyle 1}{\displaystyle \rho_{\text{por}}}
\frac{\displaystyle 1}{\displaystyle 3 \Phi }
\accentset{\text{por}}{\mathbf{T}}_{\text{ep}} : \accentset{\text{por}}{\mathbf{C}}_{\text{ep}}
- \frac{\displaystyle \partial \psi(\hat{\mathbf C}_{\text{e}}, \check{\mathbf C}_{\text{ie}}, s_{\text{e}},
\Phi)}{\displaystyle \partial \Phi} \Big) \dot{\Phi}
+ \Big( \frac{1}{2 \rho_{\scriptscriptstyle \text{por}}} \hat{\mathbf S}_{\text{ep}} -
\frac{\displaystyle \partial \psi_{\text{el}}(\hat{\mathbf{C}}_{\text{e}}, \Phi)}
{\displaystyle \partial \hat{\mathbf{C}}_{\text{e}}} \Big):
\stackrel{\displaystyle \cdot} {\hat{\mathbf{C}}}_{\text{e}} \\ +
\Big(\frac{1}{2 \rho_{\scriptscriptstyle \text{por}}} \check{\mathbf X} -
\frac{\displaystyle \partial \psi_{\text{kin}}(\check{\mathbf{C}}_{\text{ie}}, \Phi)}
{\displaystyle \partial \check{\mathbf{C}}_{\text{ie}}} \Big)
: \stackrel{\displaystyle \cdot} {\check{\mathbf{C}}}_{\text{ie}} +
\frac{1}{\rho_{\scriptscriptstyle \text{por}}} \big(\hat{\mathbf{C}}_{\text{e}} \hat{\mathbf S}_{\text{ep}} -
\hat{\mathbf X} \big) :  \hat{\mathbf{D}}_{\text{i}} +
\frac{1}{\rho_{\scriptscriptstyle \text{por}}} \big(\check{\mathbf{C}}_{\text{ie}} \check{\mathbf X} \big)
:\check{\mathbf{D}}_{\text{ii}} \\
 - \frac{\displaystyle \partial \psi_{\text{iso}} (s_{\text{e}}, \Phi)}{\displaystyle \partial s_{\text{e}}}  \dot{s}_{\text{e}}.
\end{multline}
Taking the potential relations \eqref{hyperelastic} into account and recalling the
abbreviations \eqref{MandelLikeBackstress}, \eqref{EffectiveStress}, the internal dissipation
takes the following simple form
\begin{equation}\label{dissi2}
\delta_{\text{i}} =  \Big( \frac{\displaystyle 1}{\displaystyle \rho_{\text{por}}}
\frac{\displaystyle 1}{\displaystyle 3 \Phi }
\accentset{\text{por}}{\mathbf{T}}_{\text{ep}} : \accentset{\text{por}}{\mathbf{C}}_{\text{ep}}
- \frac{\displaystyle \partial \psi(\hat{\mathbf C}_{\text{e}}, \check{\mathbf C}_{\text{ie}}, s_{\text{e}},
\Phi)}{\displaystyle \partial \Phi} \Big) \dot{\Phi}  +
\frac{1}{\rho_{\scriptscriptstyle \text{por}}} \big( \hat{\mathbf{\Sigma}} :  \hat{\mathbf{D}}_{\text{i}} +
 \check{\mathbf \Xi} :\check{\mathbf{D}}_{\text{ii}}
 -  R  \dot{s}_{\text{e}} \big).
\end{equation}
Damage-induced energy dissipation is given by the first term on the right-hand side of \eqref{dissi2}.
The remaining part corresponds to the energy dissipation induced by isochoric plasticity.
In other words, this relation encompasses the concept that both plasticity and damage are dissipative processes.
As a sufficient condition for the Clausius-Duhem inequality,
we assume that both contributions are non-negative
\begin{equation}\label{dissi3}
\delta^{\text{damage}}_{\text{i}} :=  \Big( \frac{\displaystyle 1}{\displaystyle \rho_{\text{por}}}
\frac{\displaystyle 1}{\displaystyle 3 \Phi }
\accentset{\text{por}}{\mathbf{T}}_{\text{ep}} : \accentset{\text{por}}{\mathbf{C}}_{\text{ep}}
- \frac{\displaystyle \partial \psi(\hat{\mathbf C}_{\text{e}}, \check{\mathbf C}_{\text{ie}}, s_{\text{e}},
\Phi)}{\displaystyle \partial \Phi} \Big) \dot{\Phi} \geq 0,
\end{equation}
\begin{equation}\label{dissi4}
\delta^{\text{plasticity}}_{\text{i}} :=
\frac{1}{\rho_{\scriptscriptstyle \text{por}}} \big(
(\hat{\mathbf{\Sigma}} :  \hat{\mathbf{D}}_{\text{i}} -  R  \dot{s}) +
\check{\mathbf \Xi} :\check{\mathbf{D}}_{\text{ii}}
+ R  \dot{s}_{\text{d}} \big) \geq 0.
\end{equation}
Note that the terms $\hat{\mathbf{\Sigma}} : \hat{\mathbf{D}}_{\text{i}}$ and
$\check{\mathbf \Xi} :\check{\mathbf{D}}_{\text{ii}}$ are related to the energy dissipation
in the friction element and the rate-independent dashpot, shown in Fig. \ref{fig1}.

\subsection{Yield condition and evolution equations}

Ductile damage influences both
the plastic flow and the size of the
elastic domain in the stress space.
To account for these effects, we consider the yield function as follows
\begin{equation}\label{yieldF}
f:= \| \hat{\mathbf \Sigma}^{\text{D}} \|- \sqrt{\frac{2}{3}} \big[K(\Phi) + R \big],
\end{equation}
where $K(\Phi) \geq 0$ stands for the damage-dependent yield stress.
As a sufficient condition for the inequality \eqref{dissi4}, we postulate the following
evolution equations for the inelastic flow and the inelastic flow of the substructure
\begin{equation}\label{evol}
\hat{\mathbf{D}}_{\text{i}} = \lambda_{\text{i}}
\frac{\displaystyle \hat{\mathbf \Sigma}^{\text{D}} } {\displaystyle
\| \hat{\mathbf \Sigma}^{\text{D}} \|}, \quad
\check{\mathbf{D}}_{\text{ii}} = \lambda_{\text{i}} \ \varkappa(\Phi) \ \check{\mathbf \Xi}^{\text{D}},
\end{equation}
\begin{equation}\label{evol2}
\dot{s} = \sqrt{\frac{2}{3}} \lambda_{\text{i}}, \quad
\dot{s}_{\text{d}} = \frac{\beta(\Phi)}{\gamma(\Phi)} \dot{s} R,
\end{equation}
where $\varkappa(\Phi) \geq 0$ and $\beta(\Phi) \geq 0$ are damage-dependent hardening parameters and
$\lambda_{\text{i}} \geq 0$ is the inelastic multiplier.
It was shown by \cite{ShutovLandgraf} (Section 2.2) that the flow rule $\eqref{evol}_1$
corresponds to the flow rule previously formulated by \cite{SimoMiehe1992} on the current configuration.
It follows from $\eqref{evol}_1$ that
$\lambda_{\text{i}} = \| \hat{\mathbf{D}}_{\text{i}} \|$. Thus, $\lambda_{\text{i}}$
 corresponds to the intensity of the inelastic flow. In this paper we describe it
based on Peryzna's rate-dependent overstress theory \citep{Perzyna1966}
\begin{equation}\label{perz}
\lambda_{\text{i}} = \frac{\displaystyle 1}{\displaystyle
\eta}\Big\langle \frac{\displaystyle f}{\displaystyle f_0}
\Big\rangle^{m}, \quad
\langle x \rangle := \text{max}(x,0).
\end{equation}
Here, $f_0 >0$ is used to obtain a non-dimensional term in the
angle bracket.\footnote{Thus, $f_0$  is not a material parameter.}
For simplicity, we assume that the viscosity parameters
$\eta \geq 0$ and $m \geq 1$ are not affected by damage.

The evolution equation $\eqref{evol}_1$ complies with the normality flow
rule.\footnote{We postulate that
the porous material obeys the normality rule as soon as the normality
rule holds true for the matrix material \citep{Berg1969}.}
Both inelastic flows governed by \eqref{evol} are incompressible. Indeed, after some computations we arrive at
\begin{equation}\label{inco2}
 \frac{d}{d t}(\text{det} \accentset{\text{por}}{\mathbf{F}}_{\text{i}}) =
 (\text{det} \accentset{\text{por}}{\mathbf{F}}_{\text{i}}) \
 \text{tr} \big(  \hat{\mathbf{D}}_{\text{i}}  \big) \stackrel{\eqref{evol}_1}{=} 0,
 \
 \frac{d}{d t}(\text{det} \accentset{\text{por}}{\mathbf{F}}_{\text{ii}}) = (\text{det} \accentset{\text{por}}{\mathbf{F}}_{\text{ii}}) \
 \text{tr} \big(  \check{\mathbf{D}}_{\text{ii}}  \big) \stackrel{\eqref{evol}_2}{=}0.
\end{equation}
Therefore, under appropriate initial conditions we have
\begin{equation}\label{inco}
\text{det} \accentset{\text{por}}{\mathbf F}_{\text{i}} = \text{det} \accentset{\text{por}}{\mathbf F}_{\text{ii}} =
\text{det} \accentset{\text{por}}{\mathbf C}_{\text{i}} = \text{det} \accentset{\text{por}}{\mathbf C}_{\text{ii}}=1.
\end{equation}
In particular, the volume changes are accommodated by
the elastic bulk strain and the damage-induced expansion of the material, but not by the
inelastic deformation. Thus, $\det \mathbf F = \det \hat{\mathbf F}_{\text{e}} \ \Phi$.

Note that in case of constant $\gamma(\Phi)$ and $\beta(\Phi)$,
the classical Voce rule of isotropic hardening
is reproduced by $\eqref{evol2}_2$. In that case, the parameter $\beta$ governs the saturation
rate. For simplicity, we may postulate $\beta(\Phi) = \beta_0 = const$.
A similar assumption for the kinematic hardening would imply
\begin{equation}\label{KinematHardeningDegrad}
c(\Phi) \varkappa(\Phi) = const = c_0 \varkappa_0, \quad \varkappa(\Phi)
\stackrel{\eqref{DegradationHardening}_1}{=} \varkappa_0 \exp \big(\text{KRR} \cdot (\Phi-1)\big).
\end{equation}
Moreover, we note that the yield stress $K(\Phi)$ is connected with the
isotropic hardening of the material, since it appears in combination with $R$ (cf. \eqref{yieldF}).
Therefore, it is natural to assume that $K(\Phi)$ is affected by damage in
the same way as the isotropic hardening. More precisely,
we assume\footnote{This relation is reasonable
for small pore volume fractions, which we assume in the current study.
Numerical simulations performed by \cite{Fritzen2012} demonstrate that
none of the classical yield conditions can predict the material yielding
at pore volume fraction above 10\%, and
a refined constitutive modeling is needed.}
\begin{equation}\label{YieldStressDegrad}
K(\Phi) = K_0 \exp \big(-\text{IRR} \cdot (\Phi-1)\big).
\end{equation}

Within our framework, the damage evolution is postulated as a function of the inelastic strain
rate $\hat{\mathbf{D}}_{\text{i}}$, the effective stress $\hat{\mathbf \Sigma}$,
the backstress $\hat{\mathbf X}$, and the damage variable $\Phi$
\begin{equation}\label{DamageEvolution}
\dot{\Phi} = \dot{\Phi} \big( \hat{\mathbf{D}}_{\text{i}}, \hat{\mathbf \Sigma}, \hat{\mathbf X}, \Phi \big).
\end{equation}
We consider ductile damage as a strain-driven process. Therefore, we suppose that $\dot{\Phi}$
is a homogeneous function of $\hat{\mathbf{D}}_{\text{i}}$
\begin{equation}\label{HomogFunction}
\dot{\Phi} \big( \alpha \hat{\mathbf{D}}_{\text{i}}, \hat{\mathbf \Sigma}, \hat{\mathbf X}, \Phi \big) =
 \alpha \dot{\Phi} \big(\hat{\mathbf{D}}_{\text{i}}, \hat{\mathbf \Sigma}, \hat{\mathbf X}, \Phi \big),
\quad \text{for all} \ \alpha \geq 0.
\end{equation}
It follows immediately from this relation that the damage does not evolve if the plastic flow is frozen.
 Next, for simplicity, we do not consider any material curing in this study. In combination with
the inequality \eqref{dissi3} we have the following restrictions
\begin{equation}\label{RestricOnPhi}
\dot{\Phi} \geq 0, \quad \dot{\Phi} = 0 \ \text{for} \
\frac{\displaystyle 1}{\displaystyle \rho_{\text{por}}}
\frac{\displaystyle 1}{\displaystyle 3 \Phi }
\accentset{\text{por}}{\mathbf{T}}_{\text{ep}} : \accentset{\text{por}}{\mathbf{C}}_{\text{ep}}
< \frac{\displaystyle \partial \psi(\hat{\mathbf C}_{\text{e}}, \check{\mathbf C}_{\text{ie}}, s_{\text{e}},
\Phi)}{\displaystyle \partial \Phi}.
\end{equation}
Taking into account that $\frac{\displaystyle \partial \psi(\hat{\mathbf C}_{\text{e}}, \check{\mathbf C}_{\text{ie}}, s_{\text{e}},
\Phi)}{\displaystyle \partial \Phi} \leq 0$ (cf. Section 2.3), the restriction $\eqref{RestricOnPhi}_2$ states
that the damage-induced expansion is impossible whenever the hydrostatic pressure exceeds a
certain limit.
As it was pointed out by \cite{BaoWierzbicki2005}, fracture never occurs in
some materials under very high hydrostatic pressure. It is impressive that a similar restriction is obtained here
in a natural way as a sufficient condition for thermodynamic consistency.

Although the presented framework is based on phenomenological assumptions, the
damage evolution law \eqref{DamageEvolution} provides an entry point for
micromechanical approaches to damage modeling (cf. the comprehensive review by \cite{Besson2010}).
It may include the modeling of void nucleation, growth, and coalescence
(some fundamental approaches are discussed by \cite{Marini1985, PardoenHutchinson2000, Xue2007, Bruenig2014}).
Since the validation and calibration of such models requires certain
microstructural observations, these models
may be helpful \emph{in extending the applicability range} of the purely
phenomenological models.
On the other hand, as emphasized by \cite{HammiHorstemeyer2007},
the calibration of physics-based \emph{anisotropic} damage models is highly non-trivial:
Taking the directional character of data for anisotropic analysis into account, some quantitative experimental data like void growth and coalescence are hardly measurable.

There is an abundance of literature dealing with specific fracture criteria
(see, for example, \cite{KhanLiu2012} and references therein).
Therefore this issue is not addressed in the present publication.

\subsection{Void nucleation rule}

\begin{figure}\centering
\psfrag{A}[m][][1][0]{$\sigma_{1 1}$}
\psfrag{B}[m][][1][0]{$\sqrt{3} \sigma_{1 2}$}
\psfrag{C}[m][][1][0]{$\frac{\displaystyle \dot{N}_{\text{tens}}}{\displaystyle n_{\text{tens}} \ \lambda_{\text{i}}}$}
\psfrag{D}[m][][1][0]{$\frac{\displaystyle \dot{N}_{\text{tens}}}{\displaystyle n_{\text{tens}} \ \lambda_{\text{i}}}$}
\psfrag{E}[m][][1][0]{$\frac{\displaystyle \dot{N}_{\text{shear}}}{\displaystyle n_{\text{shear}} \ \lambda_{\text{i}}}$}
\psfrag{F}[m][][1][0]{$\frac{\displaystyle \dot{N}_{\text{shear}}}{\displaystyle n_{\text{shear}} \ \lambda_{\text{i}}}$}
\psfrag{G}[m][][1][0]{$K_{\text{tens}}=0.98$}
\psfrag{N}[m][][1][0]{$K_{\text{tens}}=0.90$}
\psfrag{H}[m][][1][0]{$K_{\text{shear}}=0.98$}
\psfrag{M}[m][][1][0]{$K_{\text{shear}}=0.90$}
\scalebox{0.85}{\includegraphics{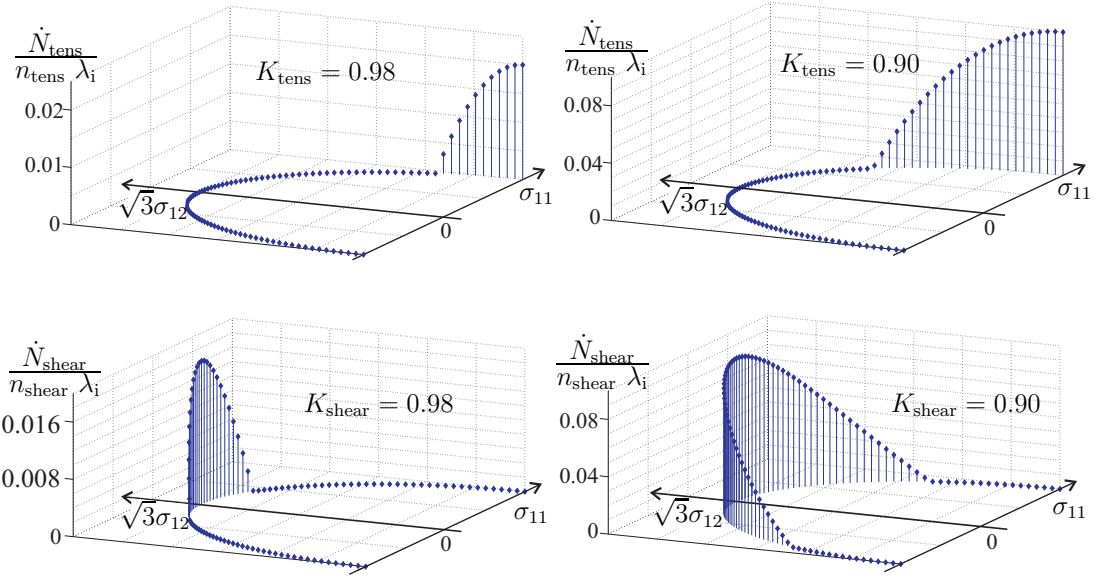}} \caption{Distribution of void nucleation rates in the $(\sigma_{1 1}, \sqrt{3} \sigma_{1 2})$-space
where $\hat{\mathbf \Sigma} =
\sigma_{1 1} \mathbf{e}_1 \otimes \mathbf{e}_1 + \sigma_{1 2} ( \mathbf{e}_1 \otimes \mathbf{e}_2 + \mathbf{e}_2 \otimes \mathbf{e}_1 )$.
Top: void nucleation rate under tension $\dot{N}_{\text{tens}} / (n_{\text{tens}} \ \lambda_{\text{i}})$ (cf. \eqref{EvolUndTension}).
Bottom: void nucleation rate under shear $\dot{N}_{\text{shear}} / (n_{\text{shear}} \ \lambda_{\text{i}})$ (cf. \eqref{EvolUndShear}).
\label{fig3}}
\end{figure}

Various void nucleation rules are available in the literature (cf. Appendix A).
In this subsection we construct a new simple rule of void nucleation
for porous ductile metals with second phases.
Let $N$ be the number of voids per unit volume of reference configuration.
The nucleation rule is based on consideration
of different nucleation mechanisms. We assume
\begin{equation}\label{VoidNucleatRule}
\dot{N} = \dot{N}_{\text{tens}} + \dot{N}_{\text{shear}} + \dot{N}_{\text{comp}},
\end{equation}
where $\dot{N}_{\text{tens}}$ and $\dot{N}_{\text{shear}}$ stand for the nucleation rates
due to separation of the matrix material under local tension and shear, respectively; $\dot{N}_{\text{comp}}$
corresponds to the crushing of inclusions under local hydrostatic compression.
It is assumed that void nucleation is driven by the inelastic flow (cf. \cite{Gurson1977, GoodsBrown1979}). Since the flow is governed by
the effective stress tensor $\hat{\mathbf \Sigma}$, a similar
assumption is made for void nucleation, which is
now controlled by the effective stress, and not by the total stress.
Let $\sigma_1$, $\sigma_2$, and $\sigma_3$
be the eigenvalues of the effective stress $\hat{\mathbf \Sigma}$.
Additionally, we introduce the norm of its deviatoric part
\begin{equation}\label{NormDeviator}
\mathfrak{F} : =  \|  \hat{\mathbf \Sigma}^{\text{D}} \|.
\end{equation}
We postulate the void nucleation rate under local tension as follows
\begin{equation}\label{EvolUndTension}
\dot{N}_{\text{tens}} = n_{\text{tens}} \
\lambda_{\text{i}} \Big\{ \Big \langle \frac{\displaystyle \sigma_1}{\displaystyle \sqrt{3/2} \mathfrak F} - K_{\text{tens}} \Big \rangle +
\Big \langle \frac{\displaystyle \sigma_2}{\displaystyle \sqrt{3/2} \mathfrak F} - K_{\text{tens}} \Big \rangle +
\Big \langle \frac{\displaystyle \sigma_3}{\displaystyle \sqrt{3/2} \mathfrak F} - K_{\text{tens}} \Big \rangle \Big\}.
\end{equation}
Here, the parameter $n_{\text{tens}} \geq 0$ controls the intensity of the void nucleation, $K_{\text{tens}} < 1$ represents a certain threshold
and $\langle x \rangle = \max(x,0)$.
The distribution of the void nucleation rate $\dot{N}_{\text{tens}}$ in the effective stress space is depicted in Fig. \ref{fig3} for
two different values of $K_{\text{tens}}$. As $K_{\text{tens}} \rightarrow 1$, only the effective stress
states close to uniaxial tension contribute to void nucleation.
For smaller values of $K_{\text{tens}}$, the distribution becomes more uniform.
Next, the void nucleation rate under local shear is given by
\begin{equation}\label{EvolUndShear}
\dot{N}_{\text{shear}} = n_{\text{shear}} \
\lambda_{\text{i}} \Big\{ \Big \langle \frac{\displaystyle | \sigma_1 - \sigma_2 |}{\displaystyle \sqrt{2} \mathfrak F} - K_{\text{shear}} \Big \rangle +
\Big \langle  \frac{\displaystyle | \sigma_2 - \sigma_3 |}{\displaystyle \sqrt{2} \mathfrak F} - K_{\text{shear}} \Big \rangle +
\Big \langle  \frac{\displaystyle | \sigma_1 - \sigma_3 |}{\displaystyle \sqrt{2} \mathfrak F} - K_{\text{shear}} \Big \rangle\Big\}.
\end{equation}
Here, $n_{\text{shear}} \geq 0$ and $K_{\text{shear}} <1$. Analogously to \eqref{EvolUndTension}, the threshold
$K_{\text{shear}}$ is used to control the distribution of the nucleation rate in the stress space.
As $K_{\text{shear}} \rightarrow 1$, a distinct single peak near pure shear state is observed (cf. Fig. \ref{fig3}).
For $K_{\text{shear}} < 1$, the distribution becomes more uniform.
Finally, the void nucleation rate under compression is given by
\begin{equation}\label{EvolUndCompress}
\dot{N}_{\text{comp}} = n_{\text{comp}} \
\lambda_{\text{i}} \langle - \text{tr} (\hat{\mathbf \Sigma}) - K_\text{comp} \rangle,
\end{equation}
where $n_{\text{comp}} \geq 0$ and $K_\text{comp}$ are fixed material parameters. Observe that this type of void nucleation
depends solely on the hydrostatic pressure $- \text{tr} (\hat{\mathbf \Sigma}) /3$ and the inelastic strain rate $\lambda_{\text{i}}$.
This mechanism contributes to the overall void nucleation only if the pressure exceeds the threshold $K_\text{comp}/3$.
Note that $\dot{N}_{\text{tens}}, \dot{N}_{\text{shear}}, \dot{N}_{\text{comp}} \geq 0$ for $n_{\text{tens}}, n_{\text{shear}}, n_{\text{comp}} \geq 0$, which
forbids any void healing.

The introduced nucleation rule is compared to some classical rules in Appendix B in a qualitative way.
Moreover, a concrete numerical example will be presented in Section 4.

\subsection{Damage evolution equation}

In this paper, for simplicity, we neglect the effects of void
coalescence and suppose that the damage evolution stems solely
from the void nucleation and growth
\begin{equation}\label{DamageEvolutRule0}
\dot{\Phi} = \dot{\Phi}_{\text{nucleation}} + \dot{\Phi}_{\text{growth}}.
\end{equation}
The material expansion due to void nucleation is described by
\begin{equation}\label{DamageEvolutRule}
\dot{\Phi}_{\text{nucleation}} = v_\text{tens} \ \dot{N}_{\text{tens}} +
v_\text{shear} \ \dot{N}_{\text{shear}} + v_\text{comp} \ \dot{N}_{\text{comp}},
\end{equation}
where the constant parameters $v_\text{tens}$, $v_\text{shear}$, and $v_\text{comp}$ stand for the characteristic void volumes.
Three different parameters are used in \eqref{DamageEvolutRule}
to account for the effects of different void volumes, depending on the nucleation mechanism.
Next, one of the simplest expressions for the material expansion due to the dilatational
void growth is given by (cf. \cite{RiceTr1969})
\begin{equation}\label{VoidGrowth}
\dot{\Phi}_{\text{growth}} = d_{\text{growth}} \ ( \Phi - \Phi_0 ) \lambda_{\text{i}}
\exp \Big(\sqrt{\frac{3}{2}} \frac{\text{tr} \hat{\mathbf \Sigma} }{\mathfrak{F}} \Big),
\quad d_{\text{growth}}=\text{const} \geq 0, \ \Phi_0 =\text{const} > 0.
\end{equation}
Here, the parameter $\Phi_0$ is adopted to capture the fraction of the pre-existing pores, such that
the special case $\Phi = \Phi_0$ corresponds to the state without voids.

\subsection{Chronological summary of the main assumptions}

In Table \ref{tab1} we summarize the main modeling assumptions
and ideas which are utilized in the current study. Effects with geometric nonlinearities are marked by ${}^*$.

\begin{table}[h]
\caption{Chronological summary of the main assumptions.}
\begin{tabular}{| l |  l|}
\hline
Effects/phenomena                &  Modeling assumptions and their historical origins        \\ \hline \hline
Elasto-plastic kinematics ${}^*$   &  Multiplicative decomposition \eqref{2ndDecomp} \citep{Bilby1957}     \\ \hline
Damage-elasticity coupling         &  Isotropic elasticity; reduced $k$ and $\mu$, cf. \eqref{DegradationElasticity} \citep{Bristow1960} \\ \hline
Viscosity                        &  Peryzna's overstress theory, cf. \eqref{perz} \citep{Perzyna1966}        \\ \hline
Void growth                      &  Strain-driven void growth, cf. \eqref{VoidGrowth}  \citep{RiceTr1969}  \\ \hline
Inelastic flow with damage       &  Normality rule for damaged material  \citep{Berg1969}  \\ \hline
Elasticity ${}^*$                &  Hyperelasticity in elastic range, cf. \eqref{hyperelastic} \citep{SimoOrtiz1985} \\ \hline
Ductile damage                   &  Porosity as an adequate measure of damage     \citep{Tvergaard1990}               \\ \hline
Damage-induced expansion ${}^*$  &  Multiplicative decomposition  \eqref{1stDecomp}  \citep{BammannAifantis}    \\ \hline
Inelastic flow ${}^*$            &  Flow rules of Simo-Miehe type, cf. \eqref{evol} \citep{SimoMiehe1992}    \\ \hline
Nonlin. kinematic hardening ${}^*$  &  Multiplicative decomposition  \eqref{3rdDecomp}  \citep{Lion2000}    \\ \hline
Void nucleation                     &  New nucleation rule presented in Section 2.6                      \\ \hline
\end{tabular}
\label{tab1}
\end{table}



\section{Numerics}

\subsection{Transformation of equations to the porous configuration}

In order to simplify the numerical treatment of the constitutive equations, they
will be transformed to the porous configuration $\accentset{\text{por}}{\mathcal{K}}$.
First, recalling \eqref{2ndDecomp} and \eqref{RCGT1}, we note that
\begin{equation}\label{InvarianceOfInvariants}
\text{tr} \Big( \hat{\mathbf{C}}^n_{\text{e}} \Big) = \text{tr} \Big(
\big(\accentset{\text{por}}{\mathbf{C}}_{\text{ep}} \ \accentset{\text{por}}{\mathbf{C}}^{-1}_{\text{i}} \big)^n \Big), \quad n=1,2,3.
\end{equation}
Analogously, using \eqref{3rdDecomp} and \eqref{RCGT122}, we have
\begin{equation}\label{InvarianceOfInvariants2}
\text{tr} \Big( \check{\mathbf{C}}^n_{\text{ie}} \Big) = \text{tr} \Big( \big(
 \accentset{\text{por}}{\mathbf{C}}_{\text{i}}
 \ \accentset{\text{por}}{\mathbf{C}}^{-1}_{\text{ii}} \big)^n \Big), \quad n=1,2,3.
\end{equation}
Therefore, due to the isotropy, the free energy \eqref{freeen} can be represented in the form
\begin{equation}\label{TransformedFreeEnergy}
\psi =  \psi_{\text{el}} \big(\accentset{\text{por}}{\mathbf{C}}_{\text{ep}} \
\accentset{\text{por}}{\mathbf{C}}^{-1}_{\text{i}}, \Phi \big) + \psi_{\text{kin}} \big(\accentset{\text{por}}{\mathbf{C}}_{\text{i}}
 \ \accentset{\text{por}}{\mathbf{C}}^{-1}_{\text{ii}}, \Phi \big) +
\psi_{\text{iso}}(s_{\text{e}}, \Phi).
\end{equation}
Moreover, note that for any smooth scalar-valued function $\alpha$ we have
\begin{equation}\label{change}
\mathbf{A}^{\text{T}} \frac{\displaystyle \partial
\alpha(\mathbf{A} \mathbf{B} \mathbf{A}^{\text{T}})}{\displaystyle \partial
( \mathbf{A} \mathbf{B} \mathbf{A}^{\text{T}} )} \mathbf{A} =
\frac{\displaystyle \partial \alpha(\mathbf{A} \mathbf{B}
\mathbf{A}^{\text{T}})}{\displaystyle \partial \mathbf{B}}
\big|_{\mathbf{A} =\text{const}}.
\end{equation}
Applying pull-back transformations to the potential relations $\eqref{hyperelastic}_1$, $\eqref{hyperelastic}_2$,
and taking \eqref{TransformedFreeEnergy}, \eqref{change} into account, we obtain
\begin{equation}\label{PullBackOfStresses}
\accentset{\text{por}}{\mathbf{T}}_{\text{ep}}  =
2 \rho_{\scriptscriptstyle \text{por}}
\frac{\displaystyle \partial \psi_{\text{el}} \big(\accentset{\text{por}}{\mathbf{C}}_{\text{ep}} \
\accentset{\text{por}}{\mathbf{C}}^{-1}_{\text{i}}, \Phi \big)}
{\displaystyle \partial \accentset{\text{por}}{\mathbf{C}}_{\text{ep}} }\big|_{\
\accentset{\text{por}}{\mathbf{C}}_{\text{i}} = \text{const}}, \
\accentset{\text{por}}{\mathbf{X}}  =
2 \rho_{\scriptscriptstyle \text{por}}
\frac{\displaystyle \psi_{\text{kin}} \big(\accentset{\text{por}}{\mathbf{C}}_{\text{i}}
 \ \accentset{\text{por}}{\mathbf{C}}^{-1}_{\text{ii}}, \Phi \big)}
{\displaystyle \partial  \accentset{\text{por}}{\mathbf{C}}_{\text{i}}  }\big|_{\ \accentset{\text{por}}{\mathbf{C}}_{\text{ii}} = \text{const}}.
\end{equation}
If the concrete ansatz \eqref{spec1}, \eqref{spec2} is adopted, then the stresses are given by
\begin{equation}\label{trans41}
\accentset{\text{por}}{\mathbf{T}}_{\text{ep}} = \Phi^{-1} \big( k(\Phi) \ \text{ln}\sqrt{\text{det} (\accentset{\text{por}}{\mathbf{C}}_{\text{ep}})} \
\accentset{\text{por}}{\mathbf{C}}_{\text{ep}}^{-1} + \mu(\Phi) \ \accentset{\text{por}}{\mathbf{C}}_{\text{ep}}^{-1} (
\overline{\accentset{\text{por}}{\mathbf{C}}}_{\text{ep}} \accentset{\text{por}}{\mathbf{C}}^{-1}_{\text{i}} )^{\text{D}} \big).
\end{equation}
\begin{equation}\label{trans42}
\accentset{\text{por}}{\mathbf{X}} = \Phi^{-1} \frac{c(\Phi)}{2} \ \accentset{\text{por}}{\mathbf{C}}_{\text{i}}^{-1}
(\accentset{\text{por}}{\mathbf{C}}_{\text{i}}  \accentset{\text{por}}{\mathbf{C}}_{\text{ii}}^{-1})^{\text{D}}.
\end{equation}
Substituting \eqref{trans41} into \eqref{RelatBet2ndPK}, for the 2nd Piola-Kirchhoff stress we obtain
\begin{equation}\label{trans43}
\tilde{\mathbf T} = k(\Phi) \ \big( \text{ln}\sqrt{\text{det} (\mathbf C)} - \text{ln}\Phi  \big) \
\mathbf C^{-1} + \mu(\Phi) \ \mathbf C^{-1} (\overline{\mathbf C}  \accentset{\text{por}}{\mathbf{C}}_{\text{i}}^{-1})^{\text{D}}.
\end{equation}
Next, we consider the invariants (moments) of the effective stress tensor
\begin{equation}\label{trace}
\text{tr} \big( \hat{\mathbf \Sigma}^n \big)
\stackrel{\eqref{EffectiveStress}}{=}
\text{tr} \big( (\hat{\mathbf{C}}_{\text{e}} \hat{\mathbf{S}}_{\text{ep}} - \hat{\mathbf{X}})^n \big) =
\text{tr} \big( (\accentset{\text{por}}{\mathbf{C}}_{\text{ep}}
\accentset{\text{por}}{\mathbf{T}}_{\text{ep}} - \accentset{\text{por}}{\mathbf{C}}_{\text{i}} \accentset{\text{por}}{\mathbf{X}})^n
\big), \ n=1,2,3.
\end{equation}
These relations imply that the eigenvalues of $\hat{\mathbf \Sigma}$ coincide
with the eigenvalues of $\accentset{\text{por}}{\mathbf{C}}_{\text{ep}}
\accentset{\text{por}}{\mathbf{T}}_{\text{ep}} -
\accentset{\text{por}}{\mathbf{C}}_{\text{i}} \accentset{\text{por}}{\mathbf{X}}$.
Moreover, since
\begin{equation}\label{FrobeniusOfDeviator}
\|  \hat{\mathbf \Sigma}^{\text{D}} \| = \sqrt{ \text{tr} \big[ (\hat{\mathbf \Sigma}^{\text{D}})^2 \big] },
\end{equation}
we obtain the driving force in the following form
\begin{equation}\label{FrobeniusOfDeviator}
\mathfrak{F} \stackrel{\eqref{NormDeviator}}{=} \|  \hat{\mathbf \Sigma}^{\text{D}} \|
= \sqrt{ \text{tr} \big[ \big( (\accentset{\text{por}}{\mathbf{C}}_{\text{ep}}
\accentset{\text{por}}{\mathbf{T}}_{\text{ep}} -
\accentset{\text{por}}{\mathbf{C}}_{\text{i}} \accentset{\text{por}}{\mathbf{X}})^{\text{D}})^2 \big] }.
\end{equation}
For the hydrostatic stress, we have the transformation rule as follows
\begin{equation}\label{traceHydrostat}
\text{tr} ( \hat{\mathbf{C}}_{\text{e}} \hat{\mathbf{S}}_{\text{ep}} ) =
\text{tr} (\accentset{\text{por}}{\mathbf{C}}_{\text{ep}}
\accentset{\text{por}}{\mathbf{T}}_{\text{ep}}).
\end{equation}
Finally, in order to transform the evolution equations, we note that
\begin{equation}\label{puba4}
\frac{d}{d t} \accentset{\text{por}}{\mathbf{C}}_{\text{i}} = 2 \accentset{\text{por}}{\mathbf F}^{\text{T}}_{\text{i}}
\ \hat{\mathbf{D}}_{\text{i}} \
\accentset{\text{por}}{\mathbf F}_{\text{i}}, \quad
\frac{d}{d t} \accentset{\text{por}}{\mathbf{C}}_{\text{ii}} = 2 \accentset{\text{por}}{\mathbf F}^{\text{T}}_{\text{ii}}
\ \check{\mathbf{D}}_{\text{ii}} \
\accentset{\text{por}}{\mathbf F}_{\text{ii}}.
\end{equation}
Substituting \eqref{evol} into these relations, we have
\begin{equation}\label{puba5}
\frac{d}{d t} \accentset{\text{por}}{\mathbf{C}}_{\text{i}} = 2 \frac{\lambda_{\text{i}}}{\mathfrak{F}}
\big( \accentset{\text{por}}{\mathbf{C}}_{\text{ep}}
\accentset{\text{por}}{\mathbf{T}}_{\text{ep}} -
\accentset{\text{por}}{\mathbf{C}}_{\text{i}} \accentset{\text{por}}{\mathbf{X}} \big)^{\text{D}}
\ \accentset{\text{por}}{\mathbf{C}}_{\text{i}}, \quad
\frac{d}{d t} \accentset{\text{por}}{\mathbf{C}}_{\text{ii}} =
2 \lambda_{\text{i}} \varkappa(\Phi) \big(\accentset{\text{por}}{\mathbf{C}}_{\text{i}} \accentset{\text{por}}{\mathbf{X}}
\big)^{\text{D}} \ \accentset{\text{por}}{\mathbf{C}}_{\text{ii}}.
\end{equation}
The system of constitutive equations
is summarized in Table \ref{tab2}.

\begin{table}
\caption{Summary of the material model}
\begin{tabular}{|l l|}
\hline
$\frac{d}{d t} \accentset{\text{por}}{\mathbf{C}}_{\text{i}} = 2 \frac{\displaystyle \lambda_{\text{i}}}{\displaystyle \mathfrak{F}}
\big( \accentset{\text{por}}{\mathbf{C}}_{\text{ep}}
\accentset{\text{por}}{\mathbf{T}}_{\text{ep}} -
\accentset{\text{por}}{\mathbf{C}}_{\text{i}} \accentset{\text{por}}{\mathbf{X}} \big)^{\text{D}}
\ \accentset{\text{por}}{\mathbf{C}}_{\text{i}}$,
& $\accentset{\text{por}}{\mathbf{C}}_{\text{i}}|_{t=0} = \accentset{\text{por}}{\mathbf{C}}_{\text{i}}^0$,
$\det \accentset{\text{por}}{\mathbf{C}}_{\text{i}}^0 =1$,  \\
$\frac{d}{d t} \accentset{\text{por}}{\mathbf{C}}_{\text{ii}} =
2 \lambda_{\text{i}} \varkappa(\Phi) \big(\accentset{\text{por}}{\mathbf{C}}_{\text{i}} \accentset{\text{por}}{\mathbf{X}}
\big)^{\text{D}} \ \accentset{\text{por}}{\mathbf{C}}_{\text{ii}}$, &
$\accentset{\text{por}}{\mathbf{C}}_{\text{ii}}|_{t=0} = \accentset{\text{por}}{\mathbf{C}}_{\text{ii}}^0$,
$\det \accentset{\text{por}}{\mathbf{C}}_{\text{ii}}^0 =1$,  \\
$\dot{s} = \sqrt{\frac{\displaystyle 2}{\displaystyle 3}} \lambda_{\text{i}}, \quad
\dot{s}_{\text{d}} = \frac{\displaystyle \beta(\Phi)}{\displaystyle  \gamma(\Phi)} \dot{s} R$, &
$s|_{t=0} = s^0, \ s_{\text{d}}|_{t=0} = s_{\text{d}}^0$, \\
$\accentset{\text{por}}{\mathbf{T}}_{\text{ep}}  =
2 \rho_{\scriptscriptstyle \text{por}}
\frac{\displaystyle \partial \psi_{\text{el}} \big(\accentset{\text{por}}{\mathbf{C}}_{\text{ep}} \
\accentset{\text{por}}{\mathbf{C}}^{-1}_{\text{i}}, \Phi \big)}
{\displaystyle \partial \accentset{\text{por}}{\mathbf{C}}_{\text{ep}} }
\big|_{\ \accentset{\text{por}}{\mathbf{C}}_{\text{i}} = \text{const}}$, &
$\accentset{\text{por}}{\mathbf{X}}  =
2 \rho_{\scriptscriptstyle \text{por}}
\frac{\displaystyle \psi_{\text{kin}} \big(\accentset{\text{por}}{\mathbf{C}}_{\text{i}}
 \ \accentset{\text{por}}{\mathbf{C}}^{-1}_{\text{ii}}, \Phi \big)}
{\displaystyle \partial \accentset{\text{por}}{\mathbf{C}}_{\text{i}} }\big|_{\ \accentset{\text{por}}{\mathbf{C}}_{\text{ii}} = \text{const}}$, \\
$R= \rho_{\scriptscriptstyle \text{por}} \frac{\displaystyle \partial \psi_{\text{iso}}
(s - s_{\text{d}}, \Phi)}{\displaystyle \partial s}|_{s_{\text{d}}= \text{const} }$, & \\
$\lambda_{\text{i}}= \frac{\displaystyle 1}{\displaystyle
\eta}\Big\langle \frac{\displaystyle f}{\displaystyle f_0}
\Big\rangle^{m}, \quad
f= \mathfrak{F}- \sqrt{\frac{2}{3}} \big[ K(\Phi) + R \big]$, &
$\mathfrak{F}=
\sqrt{\text{tr} \big[ \big( \mathbf C \tilde{\mathbf T} - \mathbf C_{\text{i}} \tilde{\mathbf X} \big)^{\text{D}} \big]^2 }$, \\
$\dot{\Phi} = \dot{\Phi} \big( \hat{\mathbf{D}}_{\text{i}}, \hat{\mathbf \Sigma}, \hat{\mathbf X}, \Phi \big)$, \quad
$\accentset{\text{por}}{\mathbf{C}}_{\text{ep}} = \Phi^{-2/3} \mathbf{C}$, &
$\tilde{\mathbf{T}} = \Phi^{1/3} \ \accentset{\text{por}}{\mathbf{T}}_{\text{ep}}$.\\
\hline
\end{tabular}
\label{tab2}
\end{table}

\subsection{Hybrid explicit/implicit time integration}

As a preliminary step, we specify the evolution equation $\eqref{puba5}_2$ governing the backstress saturation
for the case where
the energy storage is described by \eqref{spec2}. Substituting \eqref{trans42} into
$\eqref{puba5}_2$, we obtain
\begin{equation}\label{ConcreteEvolEquations2}
\frac{d}{d t} \accentset{\text{por}}{\mathbf{C}}_{\text{ii}} = \Phi^{-1} \lambda_{\text{i}} \varkappa(\Phi)
c(\Phi)  \big(
\accentset{\text{por}}{\mathbf{C}}_{\text{i}}  \accentset{\text{por}}{\mathbf{C}}_{\text{ii}}^{-1} \big)^{\text{D}}
\ \accentset{\text{por}}{\mathbf{C}}_{\text{ii}}.
\end{equation}
As is typical for metal viscoplasticity, the system of underlying equations is stiff.
Thus, an explicit time integration scheme would be stable only for very small time steps.
In case of large $c(\Phi)$, especially severe restrictions are imposed
on the time steps due to the stiff part in equation
\eqref{ConcreteEvolEquations2}. In this study we benefit
from the fact that this equation
is similar to the evolution equation governing the Maxwell fluid \citep{ShutovLandgraf}.
For this certain type of Maxwell fluid, there exists
an explicit update formula within the implicit time integration \citep{ShutovLandgraf}.
Equation \eqref{ConcreteEvolEquations2}
is treated using the explicit update formula, which leads to \eqref{ExplImplAlg10}.
The remaining evolution equations are discretized using the
explicit Euler forward method with subsequent correction of incompressibility.
An obvious advantage of \eqref{ExplImplAlg10} over straightforward
explicit discretization is that the solution ${}^{n+1}\accentset{\text{por}}{\mathbf{C}}_{\text{ii}}$
remains bounded even for very large time steps and large values of $c(\Phi)$.
\footnote{Moreover, \cite{SilbermannPAMM} have shown by a series of numerical tests that
the use of the update formula \eqref{ExplImplAlg10}
makes the integration procedure more robust and accurate
compared to the fully explicit procedure.}

Let us consider a typical time step from ${}^n t$ to ${}^{n+1} t$ with $\Delta t := {}^{n+1} t - {}^n t >0$.
Assume that ${}^{n+1} \mathbf{C} := \mathbf{C} ({}^{n+1} t)$ is known.
For the viscous case with $\eta >0$,
the hybrid explicit/implicit integration procedure is as follows
\begin{equation}\label{ExplImplAlg1}
\accentset{\text{por}}{\mathbf{C}}_{\text{ep}} = {}^n\Phi^{-2/3} \ {}^{n+1}\mathbf{C},
\end{equation}
\begin{equation}\label{ExplImplAlg2}
\accentset{\text{por}}{\mathbf{T}}_{\text{ep}} = {}^n \Phi^{-1} \big( k({}^n\Phi) \ \text{ln}\sqrt{\text{det} (\accentset{\text{por}}{\mathbf{C}}_{\text{ep}})} \
\accentset{\text{por}}{\mathbf{C}}_{\text{ep}}^{-1} + \mu({}^n\Phi) \ \accentset{\text{por}}{\mathbf{C}}_{\text{ep}}^{-1} (
\overline{\accentset{\text{por}}{\mathbf{C}}}_{\text{ep}} \ {}^{n} \accentset{\text{por}}{\mathbf{C}}^{-1}_{\text{i}} )^{\text{D}} \big),
\end{equation}
\begin{equation}\label{ExplImplAlg3}
\accentset{\text{por}}{\mathbf{X}} = {}^n\Phi^{-1} \frac{c({}^n \Phi)}{2} \ {}^n\accentset{\text{por}}{\mathbf{C}}_{\text{i}}^{-1}
({}^n\accentset{\text{por}}{\mathbf{C}}_{\text{i}}  {}^n\accentset{\text{por}}{\mathbf{C}}_{\text{ii}}^{-1})^{\text{D}}, \quad
R = {}^n \Phi^{-1} \gamma({}^n \Phi) ({}^n s - {}^n s_{\text{d}}),
\end{equation}
\begin{equation}\label{ExplImplAlg4}
\mathfrak{F}
= \sqrt{ \text{tr} \big[ \big( (\accentset{\text{por}}{\mathbf{C}}_{\text{ep}} \
\accentset{\text{por}}{\mathbf{T}}_{\text{ep}} -
{}^n\accentset{\text{por}}{\mathbf{C}}_{\text{i}} \ \accentset{\text{por}}{\mathbf{X}})^{\text{D}})^2 \big] },
\
f= \mathfrak{F}- \sqrt{\frac{2}{3}} \big[ K({}^n \Phi) + R \big], \
\lambda_{\text{i}}= \frac{\displaystyle 1}{\displaystyle
\eta}\Big\langle \frac{\displaystyle f}{\displaystyle f_0}
 \Big\rangle^{m},
\end{equation}
The inelastic flow is frozen for $\lambda_{\text{i}}=0$:
\begin{equation}\label{FrozenFlow}
{}^{n+1}\accentset{\text{por}}{\mathbf{C}}_{\text{i}} = {}^n\accentset{\text{por}}{\mathbf{C}}_{\text{i}},
\ {}^{n+1}\accentset{\text{por}}{\mathbf{C}}_{\text{ii}} = {}^n\accentset{\text{por}}{\mathbf{C}}_{\text{ii}},
\ {}^{n+1} s= {}^{n} s, \ {}^{n+1} s_{\text{d}}= {}^{n} s_{\text{d}}, \
{}^{n+1} \Phi= {}^{n} \Phi \
\text{for} \ \lambda_{\text{i}}=0.
\end{equation}
For $\lambda_{\text{i}} > 0$, the internal variables are updated in the following way:
\begin{equation}\label{ExplImplAlg9}
{}^{n+1}\accentset{\text{por}}{\mathbf{C}}_{\text{i}} = \overline{
{}^{n}\accentset{\text{por}}{\mathbf{C}}_{\text{i}} + 2 \Delta t
 \frac{\lambda_{\text{i}}}{\mathfrak{F}}
\big( \accentset{\text{por}}{\mathbf{C}}_{\text{ep}}
\accentset{\text{por}}{\mathbf{T}}_{\text{ep}} -
{}^{n} \accentset{\text{por}}{\mathbf{C}}_{\text{i}} \accentset{\text{por}}{\mathbf{X}} \big)^{\text{D}}
\ {}^{n} \accentset{\text{por}}{\mathbf{C}}_{\text{i}}},
\end{equation}
\begin{equation}\label{ExplImplAlg10}
{}^{n+1}\accentset{\text{por}}{\mathbf{C}}_{\text{ii}} = \overline{
{}^{n}\accentset{\text{por}}{\mathbf{C}}_{\text{ii}} + \Delta t \
{}^{n}\Phi^{-1} \lambda_{\text{i}} \ \varkappa({}^{n}\Phi) \
 c({}^{n}\Phi) \ {}^{n+1}\accentset{\text{por}}{\mathbf{C}}_{\text{i}}},
\end{equation}
\begin{equation}\label{ExplImplAlg11}
{}^{n+1}s = {}^{n}s + \Delta t \ \sqrt{\frac{\displaystyle 2}{\displaystyle 3}} \lambda_{\text{i}},
\quad
{}^{n+1}s_{\text{d}} = {}^{n}s_{\text{d}} + \Delta t \
\sqrt{\frac{\displaystyle 2}{\displaystyle 3}} \lambda_{\text{i}}
\frac{\displaystyle \beta({}^{n}\Phi)}{\displaystyle  \gamma({}^{n}\Phi)} R,
\end{equation}
\begin{equation}\label{ExplImplAlg5}
\{ \sigma_1, \sigma_2, \sigma_3  \} = \text{eigenvalues of}  \
( \accentset{\text{por}}{\mathbf{C}}_{\text{ep}} \
\accentset{\text{por}}{\mathbf{T}}_{\text{ep}} -
{}^n\accentset{\text{por}}{\mathbf{C}}_{\text{i}} \ \accentset{\text{por}}{\mathbf{X}} ), \quad
\text{tr} \hat{\mathbf \Sigma} = \text{tr} (\accentset{\text{por}}{\mathbf{C}}_{\text{ep}} \
\accentset{\text{por}}{\mathbf{T}}_{\text{ep}}),
\end{equation}
\begin{equation}\label{ExplImplAlg61}
\dot{N}_{\text{tens}} = \dot{N}_{\text{tens}} (\lambda_{\text{i}}, \sigma_1, \sigma_2, \sigma_3,  \mathfrak{F}), \quad
\dot{N}_{\text{shear}} = \dot{N}_{\text{shear}} (\lambda_{\text{i}}, \sigma_1, \sigma_2, \sigma_3,  \mathfrak{F}),
\end{equation}
\begin{equation}\label{ExplImplAlg62}
\dot{N}_{\text{comp}} = \dot{N}_{\text{comp}} (\lambda_{\text{i}}, \text{tr}\hat{\mathbf \Sigma},  \mathfrak{F}),
\end{equation}
\begin{equation}\label{ExplImplAlg7}
\dot{\Phi}_{\text{nucleation}} = v_\text{tens} \ \dot{N}_{\text{tens}} + v_\text{shear} \ \dot{N}_{\text{shear}} + v_\text{comp} \ \dot{N}_{\text{comp}},
\end{equation}
\begin{equation}\label{ExplImplAlg8}
\dot{\Phi}_{\text{growth}} = d_{\text{growth}} \   ( {}^n\Phi - \Phi_0 ) \lambda_{\text{i}} \exp \Big(\sqrt{\frac{3}{2}} \frac{\text{tr} \hat{\mathbf \Sigma} }{\mathfrak{F}} \Big),
\quad \dot{\Phi} = \dot{\Phi}_{\text{nucleation}} + \dot{\Phi}_{\text{growth}},
\end{equation}
\begin{equation}\label{ExplImplAlg12}
{}^{n+1}\Phi = {}^{n}\Phi + \Delta t \ \dot{\Phi}.
\end{equation}
Note that the update formulas \eqref{ExplImplAlg9} and \eqref{ExplImplAlg10}
exactly preserve the incompressibility conditions $\det \accentset{\text{por}}{\mathbf{C}}_{\text{i}} =1$
and $\det \accentset{\text{por}}{\mathbf{C}}_{\text{ii}} =1$
which is advantageous for the prevention of error accumulation
\citep{ShutovKr2010}

In case of small time steps,
the hybrid explicit/implicit integration scheme \eqref{ExplImplAlg1}---\eqref{ExplImplAlg12} is more efficient than
the fully implicit scheme, since the solution is given in a closed form and
no iteration procedure is required. Due to the reduced computational effort, the presented algorithm is well suited
for use within a globally explicit FEM procedure.
The algorithm is implemented into the FEM code Abaqus/Explicit
adopting the user material subroutine VUMAT. An FEM simulation of a representative boundary value problem
is presented in Section 5.

\section{Comparison with experiments}

\begin{figure}\centering
\scalebox{0.95}{\includegraphics{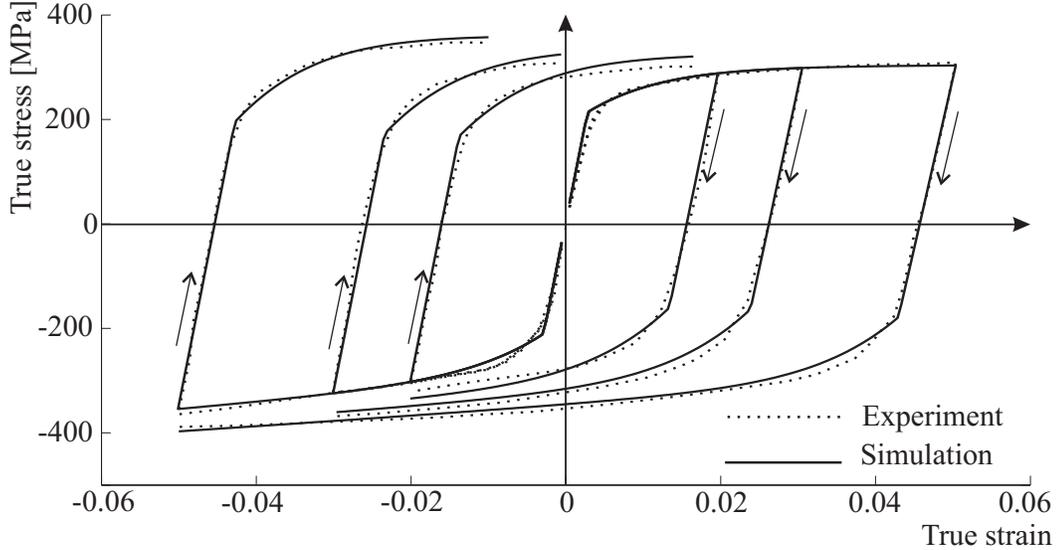}}
\caption{Uniaxial
tension-followed-by-compression
and compression-followed-by-tension
tests. Experimental
data for the cast A356 aluminium alloy \citep{Horstemeyer1998} are shown by dotted lines.
Simulation results are depicted by solid lines.
\label{fig4}}
\end{figure}

In order to validate the proposed material model,
some experimental observations of the Bauschinger effect dependent on accumulated ductile
damage are needed. In this study we use a series of quasistatic uniaxial tension-followed-by-compression
and compression-followed-by-tension tests
presented by \cite{Horstemeyer1998} for a
cast A356 aluminium alloy.\footnote{Although the ductility of this alloy is very
limited, it plays an important role in the automotive industry. Its
microstructural characterization was presented by \cite{Gall1999}.}
The flow curves pertaining to six different experiments are shown in Figure \ref{fig4} (dotted lines).
As can be seen from this figure, the initial yield stress under tension coincides with the initial
yield stress under compression. But, for the plastically deformed samples,
the true stresses under tension are essentially smaller than the corresponding stresses under compression.
We interpret this tension-compression asymmetry as a
consequence of ductile damage, since the damage evolution under tension
is typically more intense than under compression.
For the considered cast alloy, the damage effect is significant even at low tensile strains.
Moreover, a distinct Bauschinger effect
is observed in all experiments (cf. Figure \ref{fig4}).
It was noted by \cite{Jordon2007} that the Bauschinger effect
induced by tensile prestrains is less pronounced than the same effect after compression.\footnote{
The same asymmetry of Bauschinger effect was also observed
in porous sintered steels by \cite{Deng2005}.}
Such behavior indicates that the damage evolution essentially influences the kinematic hardening.

\begin{table}[h]
\caption{Material parameters.}
\begin{tabular}{| l | l | l|}
\hline
parameter & value       &  brief explanation       \\ \hline \hline
$k_0$          & 73500  MPa     &   initial bulk modulus                       \\ \hline
$\mu_0$        & 28200   MPa    &   initial shear modulus                     \\ \hline
$c_0$          & 6399.8  MPa    &   kinematic hardening parameter             \\ \hline
$\varkappa_0$          & 0.015106  MPa$^{-1}$    &  kinematic  hardening parameter    \\ \hline
$\gamma_0$     & 1442.2  MPa    &  isotropic  hardening parameter     \\ \hline
$\beta_0$       & 1.852         &  isotropic  hardening parameter     \\ \hline
$K_0$          &  210 MPa       &  initial yield stress   \\ \hline
$m$            &  1               &  viscosity parameter   \\ \hline
$\eta$            &  100 s           &  viscosity parameter   \\ \hline
KRR              &   67.63          & kinematic hardening reduction rate  \\ \hline
IRR              &   29.97          & isotropic hardening reduction rate  \\ \hline
BRR              &   45         & bulk modulus reduction rate  \\ \hline
SRR              &   30          & shear modulus reduction rate  \\ \hline
$v_\text{tens}$  & $10^{-7}$  $\text{mm}^3$   & void volume nucleated under tension  \\ \hline
$v_\text{shear}$  & $10^{-7}$  $\text{mm}^3$   & void volume nucleated under shear  \\ \hline
$n_{\text{tens}}$  & 2773000 $\text{mm}^{-3} \text{MPa}^{-1}$   & void nucleation parameter (tension)  \\ \hline
$K_{\text{tens}}$  & 0.79                          & void nucleation parameter (tension)  \\ \hline
$n_{\text{shear}}$  & 17188000 $\text{mm}^{-3} \text{MPa}^{-1}$   & void nucleation parameter (shear)  \\ \hline
$K_{\text{shear}}$  & 0.9353     & void nucleation parameter (shear)  \\ \hline
$n_{\text{comp}}$  & 0 $\text{mm}^{-3} \text{MPa}^{-1}$   & void nucleation parameter (compression)  \\ \hline
\hline
$f_0$            &  1 MPa           &  constant (not a material parameter)   \\ \hline
\end{tabular}
\label{tab3}
\end{table}

For the validation of the material model, a series of numerical tests is performed.
As the reference configuration we choose the initial configuration occupied by the body at $t=0$.
Since the initial state is stress free, we put
$\accentset{\text{por}}{\mathbf{C}}_{\text{i}}|_{t=0} = \mathbf{1}$.
Furthermore, the initial state of the cast material is assumed to be isotropic.
Therefore, we
have
$\accentset{\text{por}}{\mathbf{C}}_{\text{ii}}|_{t=0} = \mathbf{1}$.\footnote{Constitutive relations can be adopted to the initial plastic anisotropy
by an appropriate choice of the initial conditions \citep{ShutovPfeiffer}.}
The remaining initial conditions are given by
$s|_{t=0} = s_{\text{d}}|_{t=0} = 0$, $\Phi|_{t=0} = 1$.

We define the set of material parameters as follows.
Firstly, void growth is neglected here:
$d_{\text{growth}} = 0$.\footnote{Thus, we assume for simplicity that
the damage in this case is dominated by void nucleation.}
Moreover, we neglect void nucleation under compression by
putting $n_{\text{comp}} = 0$. Thus, the values of $K_{\text{comp}}$ and $v_\text{comp}$ become irrelevant.
The elasticity parameters $k_0$ and $\mu_0$ are identified
using the stress-strain curve in the elastic range.
The initial yield stress $K_0$ is estimated at a transition from elasticity to plasticity
under quasistatic loading conditions.
In general, the viscosity parameters $m$ and $\eta$ can be estimated using a series
of tests with different loading rates.
In this section we neglect the real viscous effects and adopt the parameters $m$ and $\eta$
only as numerical regularization parameters.
All simulations will be performed with a low loading rate such
that the overstress $f$ will remain below $10$ MPa.
Next, the nucleation parameters $K_{\text{tens}}$, $K_{\text{shear}}$,
$n_{\text{tens}}$, and $n_{\text{shear}}$ should be identified by counting the void nucleation sites
as a function of strain under different loading conditions \citep{Horstemeyer2000}.
In this section, $v_{\text{tens}}$ and $v_{\text{shear}}$ are roughly estimated
by assuming that each void has a volume of approximately $10^{-7} \text{mm}^3$.
In general, the parameters $v_{\text{tens}}$ and $v_{\text{shear}}$ should be identified by
measuring the volume changes induced by ductile damage.
Finally, the remaining parameters can be reliably identified using the
experimental flow curves shown in Figure \ref{fig4}.
Note that only 7 material parameters
($K_0$, $\gamma_0$, $\beta_0$, $\varkappa_0$, $c_0$, IRR, KRR)
were adjusted to fit the flow curves.
The set of material parameters which is used to describe the material response is
summarized in Table \ref{tab3}.

The stress response of the material can be described by
the proposed material model
with sufficient accuracy (cf. Figure \ref{fig4}).
Within the numerical simulation, the real Bauschinger effect is slightly underestimated,
which leads to a minor discrepancy between the simulated and the real stress responses shortly
after the secondary yielding. Following the ideas of \cite{ChabocheIJP},
this discrepancy can be reduced by introducing
additional backstress.\footnote{For a model of multiplicative type it was carried out by \cite{ShutovKuprin}.}

During the phase of 5\% prestrain under tension (within the tension-followed-by-compression test),
the theoretical value of $\Phi$ can be up to $\Phi_{0.05} = 1.00428$.
The elastic modulus
$E_{0.05}$ following 5\% tension and the initial modulus $E_0$ are computed with
\begin{equation}\label{ElastModRed}
E_{0.05} = \frac{9 k(\Phi_{0.05}) \ \mu(\Phi_{0.05})}{3 k(\Phi_{0.05}) + \mu(\Phi_{0.05})}, \quad
E_0 = \frac{9 k_0 \ \mu_0}{3 k_0 + \mu_0}.
\end{equation}
The numerical simulation yields $E_{0.05} = 0.8729 \ E_{0}$, which
corresponds to $12.7 \%$ reduction.
At the same time, the elastic modulus is unaffected by the 5\% compression.
As can be seen from Figure \ref{fig4}, the elastic unloading
is captured by the model with ample accuracy, both for tensional and compressional prestrains.

Another noteworthy aspect is the following. According to Table \ref{tab3},
KRR $\approx$ 2 IRR. This means that the deterioration of the kinematic hardening
progresses twice as fast as the deterioration of the isotropic hardening.
Thus, the concept of effective stresses combined with the strain equivalence principle
is too restrictive for the cast A356 aluminium alloy, since this concept
implies that the kinematic and isotropic hardening deteriorate
with the same rate \citep{Grammenoudis, BroeckerMatzenmiller2014}.

\begin{figure}\centering
\psfrag{A}[m][][1][0]{$N / \text{mm}^{3}$}
\psfrag{S}[m][][1][0]{$s$}
\psfrag{B}[m][][1][0]{$\# / \text{mm}^{2}$}
\scalebox{0.90}{\includegraphics{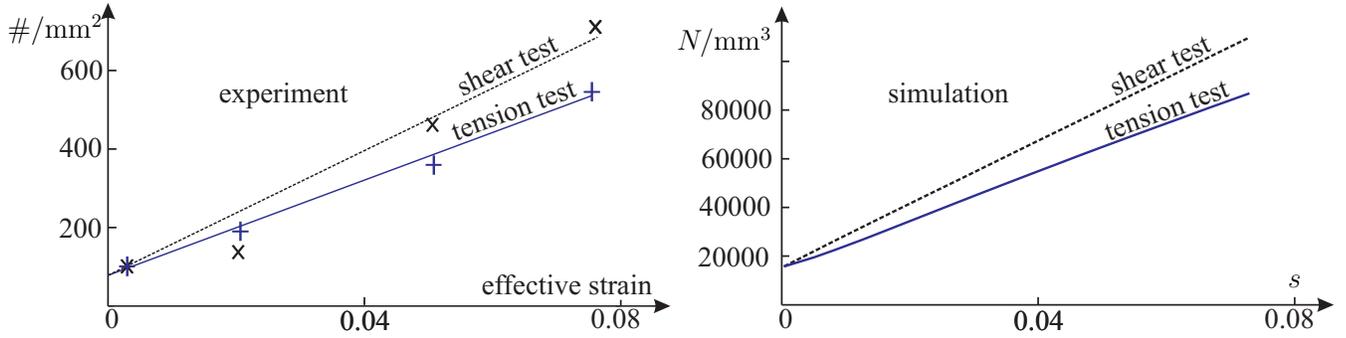}} \caption{Left: Experimental results for
the void nucleation in the cast A356 aluminium alloy \citep{Horstemeyer2000}. Number of voids
per unit area is depicted.
Right: Simulation results for the void number $N$ as a function of the accumulated
inelastic arc-length $s$. Depending on the choice of the material parameters, the void nucleation
rate under shear may exceed the nucleation rate under tension.
\label{fig5}}
\end{figure}

Uniaxial experiments are insufficient to validate the void nucleation rule proposed in the current study.
Such a validation should be based on experimental observation of void nucleation for different loading conditions.
In this study we use experimental data previously reported by \cite{Horstemeyer2000}
for the cast A356 aluminium alloy, see Figure \ref{fig5} (left).
Tension and torsion are considered here.
The simulated void number $N$ is plotted versus the accumulated inelastic arc-length $s$
in Figure \ref{fig5} (right). For the simulation of both tests, the initial value
of 15000 voids per $\text{mm}^3$ is adopted.
The simulation results correspond qualitatively to the experimental data.

\section{FEM solution of a representative boundary value problem}

\begin{figure}\centering
\scalebox{0.7}{\includegraphics{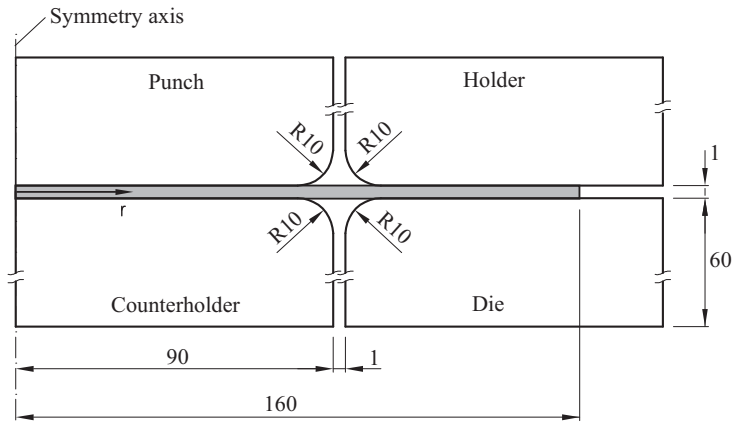}} \caption{Setup of the axisymmetric deep drawing.
\label{fig6}}
\end{figure}

An axisymmetric deep drawing of a circular sheet metal blank is considered in this section
to demonstrate the applicability of the new model and to test the stability of the
implemented numerical procedures. The setup is depicted in Fig. \ref{fig6}.
The overall forming process consists of three steps:
\begin{itemize}
\item[1] Preliminary step: Forces are gradually applied both to holder and counterholder.
\item[2] Deep drawing: The punch pushes the blank into the die.
\item[3] Spring back: The blank is released from any contact.
\end{itemize}
The material of the blank is described by the proposed
material model with parameters taken from Table \ref{tab3}, which
correspond to the A356 aluminium alloy.\footnote{
Note that the formability of cast alloys is low. For that reason,
the deep drawing process simulated here will be hardly implemented in practice. Instead, wrought alloys are used.
Nevertheless, we consider this cast alloy here as an example material because of
a rich literature dealing with its experimental characterization.
The simulation results of this subsection are used to validate the robustness
and accuracy of the numerical algorithm and to demonstrate the practicability of our approach.}
The remaining bodies are assumed to be rigid.
Coulomb friction is assumed with a friction coefficient of 0.1.
A quasistatic loading rate is chosen such that the kinetic energy is
negligible compared to the work of external forces. Mass scaling, which is
typically adopted to speed up computation, is not used here.
The FE mesh consists of 1950 elements of ABAQUS type CAX4R (4-node bilinear axisymmetric quadrilateral
element with reduced integration, hourglass control, linear geometric order). Steps 1, 2 and 3
last $0.01\text{s}$, $0.10\text{s}$ and $0.12\text{s}$, respectively. The simulation required
$460\,563$, $4\,638\,783$ and $5\,592\,654$ increments for steps 1, 2 and 3, respectively.
\begin{figure}\centering
\scalebox{0.12}{\includegraphics{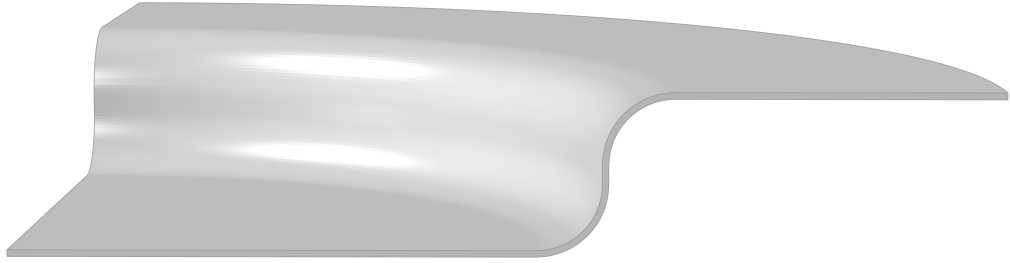}} \caption{Deformed state after deep drawing. Only a quarter of the blanket is shown.
\label{fig7}}
\end{figure}

\begin{figure}\centering
\psfrag{A}[m][][1][0]{after deep drawing}
\psfrag{B}[m][][1][0]{after spring back}
\psfrag{C}[m][][1][0]{von Mises stresses}
\scalebox{0.7}{\includegraphics{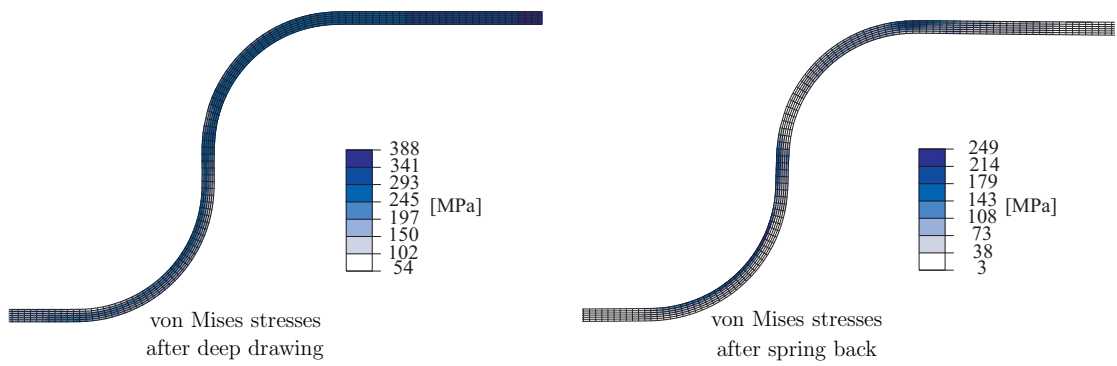}} \caption{Distribution
of von Mises stresses after deep drawing (left) and after spring back (right).
\label{fig8}}
\end{figure}

\begin{figure}\centering
\psfrag{A}[m][][1][0]{maximum principal stress}
\psfrag{B}[m][][1][0]{minimum principal stress}
\scalebox{0.7}{\includegraphics{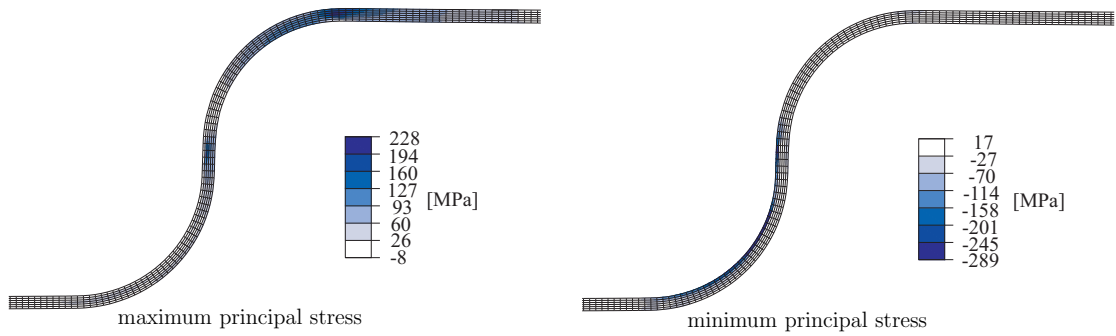}} \caption{Distribution
of maximum and minimum principal stresses after the
spring back.
\label{fig9}}
\end{figure}

It is known that the deterioration of elastic properties and the Bauschinger effect
have a strong impact on the spring back \citep{Wagoner2013}.
The deformed state after deep drawing is shown in Fig. \ref{fig7}.\footnote{An animated version of the FEM solution
can be found under http://youtu.be/YhEwI0cpIGQ and http://youtu.be/r6rY2oRUNqQ .}
Fig. \ref{fig8} shows the distribution of the von Mises equivalent stress after deep
drawing and after the spring back.
Fig. \ref{fig9} provides the distribution of the maximum and minimum principal stresses after the
spring back. Note that the residual stresses remain
high compared to the initial yield stress ($K_0=210\ \text{MPa}$).
\begin{figure}\centering
\psfrag{A}[m][][1][0]{inelastic arc length $s$}
\psfrag{B}[m][][1][0]{damage measure $\Phi-1$}
\scalebox{0.7}{\includegraphics{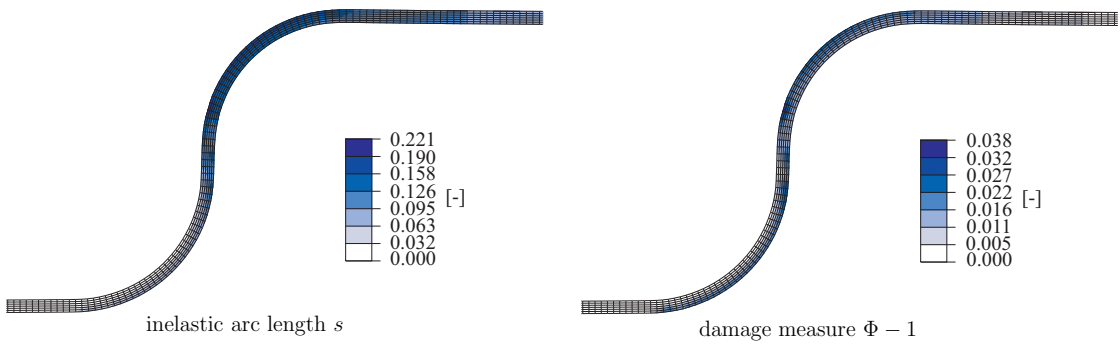}} \caption{Distribution of the
inelastic arc length $s$ (left) and damage measure $\Phi-1$ (right) after the spring back.
\label{fig10}}
\end{figure}
By comparison with the resulting inelastic arc length field $s$ (cf. Fig. \ref{fig10} (left)), it becomes clear
that highest residual stresses appear in regions with a considerable gradient of $s$.
The distribution of the damage measure $\Phi-1$ after spring back is depicted in Fig. \ref{fig10} (right).
Note that $\Phi-1$ remains in the range of a few percent and the damage is concentrated
in the deep-drawn section. The observed damage hot spots are considered as a precursor of failure.

Note that a generic local approach to ductile damage is adopted in this study.
In general, the simulation results for local damage models are pathologically mesh-dependent in case of
localizing deformation processes. Thus, the presented model should not be used for the
analysis of post-localization structural behavior.
The post-localization analysis should be
based on nonlocal damage
models which enrich the continuum formulation by introducing a physics-based
length scale (see \cite{BazantJirasek2002, Anand2012, Ahad2014, Nguyen2015} and
references therein).
Some sort of regularization is also possible using the viscosity, which
is available in the current model.
Unfortunately, this technique is effective only in a narrow range of deformation rates \citep{BazantJirasek2002}.
Still, the proposed local material model is useful
for estimating the process parameters which lead to undesired strain localization.
No strain localization was observed in the current simulation.

\section{Discussion and conclusion}

A \emph{thermodynamically consistent} approach to ductile damage and nonlinear kinematic hardening is suggested
within the framework of multiplicative plasticity.
It is well established that at the early stage of the damage
process 
the damage is adequately described by the scalar porosity variable (see, for example, \cite{Tvergaard1990}).
Thus, for simplicity, isotropic damage with no prominent orientation is assumed in this study.
To extend the current model to the case of anisotropic
damage, some physics-based tensorial damage quantities can be introduced.\footnote{Different approaches to
anisotropic damage can be found in \cite{MurakamiOhno1980, Kachanov1986, Chaboche1993, Krajcinovic1996, Bruenig2001,
Bruenig2003, Zapara2010, Zapara2012, Voyiadjis2012, Tutyshkin2014}.
}

The damage-induced \emph{volume change} is taken into account explicitly by the model kinematics for the
accurate prediction of the hydrostatic stress component.
Another feature of the proposed model is that isotropic and kinematic hardening \emph{deteriorate differently}
for increasing damage, which is essential for an accurate description of the real material response.
Moreover, the \emph{deterioration of elastic properties} is introduced for the correct prediction of residual stresses and spring back.
Next, a \emph{refined void nucleation rule}
is constructed in this study aiming at accurate description of the complex behavior
of porous ductile metals with second phases.
In particular, the new rule is suitable for materials which exhibit
a higher void nucleation rate under torsion than in case of tension.
The proposed model is validated using the experimental
observations of the void nucleation under various loading conditions
and experimental stress-strain curves. These curves carry information on the
evolution of plastic anisotropy, reduction of strain hardening capacity, and
deterioration of elastic properties.

The phenomenological formulation can be enriched by
physics-based rules for void nucleation, growth and coalescence.
The restrictions imposed on these relations by the Clausius-Duhem inequality
are \emph{obtained in an explicit form}. Interestingly, a
cut-off value for hydrostatic pressure is obtained
from irreversible thermodynamics in a natural way (cf. $\eqref{RestricOnPhi}_2$).

A relatively large number of material parameters is introduced due to a
variety of nonlinear phenomena covered by the model.
In some applications, however, certain parts of the
model can be switched off to reduce the number of parameters.
For instance, some elementary or classical void nucleation rules can be adopted (cf. Appendix A).
Alternatively, the nucleation rules \eqref{EvolUndTension} -- \eqref{EvolUndCompress}
can be calibrated using experiments (cf. \cite{Su2010} )
or lower-scale models (cf. \cite{Gall2000, DandekarShin2011}).

An efficient hybrid explicit/implicit numerical procedure is presented. This procedure exhibits
\emph{increased stability and accuracy} being compared to a purely explicit time stepping.
The feasibility of the proposed algorithm is tested numerically and the algorithm
is shown to be suitable for explicit FEM analysis.

Although the model has been developed for forming simulations,
another promising application includes
the prediction of damage and plastic anisotropy in ultra-fine
grained materials produced by severe plastic deformations
\citep{Wagner2010, Frint2011, Neugebauer2012}.
An extension of the framework to models of creep
damage mechanics \citep{Altenbach2002, Altenbach2003} 
is possible as well.

\section*{Acknowledgement}
The financial support provided by DFG within SFB 692 is acknowledged.

\section*{Appendix A (some void nucleation rules)}

In this appendix we briefly recall some of the void nucleation rules that are
available in the literature.
For simplicity, only monotonic loading is considered here.
Let $N$ be the void
number per unit volume of the reference configuration.
A very simple void nucleation rule says that $N$ is a linear function
of the accumulated plastic arc-length $s$ \citep{Gurland1972}
\begin{equation}\label{GurlandEquation}
\dot{N} = \lambda_{\text{i}} \ n_{\text{Gurland}}  \quad \stackrel{\eqref{evol2}}{\Rightarrow}
\quad
\frac{\displaystyle \partial N}{\displaystyle \partial s} = \sqrt{3/2} \ n_{\text{Gurland}}.
\end{equation}
Here, $n_{\text{Gurland}} \geq 0$ is a material parameter. In other words, the
void nucleation rate $\frac{\partial N}{\partial s}$ is assumed to be independent
of the stage of the deformation process and of the stress state.

In order to capture the dependence on the stage of the deformation process,
relation \eqref{GurlandEquation} was generalized by \cite{Gurson1977}.
Here we consider a popular modification of Gurson's rule. Let $f_{\text{void}}$ be the void volume
fraction. For the strain-controlled nucleation, \cite{ChuNeedleman1980}
assumed that the void nucleation is governed by a normal distribution law
\begin{equation}\label{ChuNeedleman1}
\dot{f}_{\text{void}} = \frac{f_{\text{N}}}{S_{\text{N}} \sqrt{2 \pi}} \exp \Big[ - \frac{\displaystyle 1}{\displaystyle 2}
\Big( \frac{s - s_{\text{N}}}{S_{\text{N}}}\Big)^2 \Big] \ \dot{s},
\end{equation}
where $f_{\text{N}}, S_{\text{N}}, s_{\text{N}} \geq 0$ are material parameters;
$s$ denotes the accumulated plastic arc-length.
Note that \eqref{GurlandEquation} can be approximated
by \eqref{ChuNeedleman1} with large $f_{\text{N}}$ and $S_{\text{N}}$ and by assuming
that $\dot{f}_{\text{void}}$ is proportional to $\dot{N}$. Another modification of Gurson's rule was proposed by \cite{ChuNeedleman1980}
for stress-controlled nucleation.

Rule \eqref{ChuNeedleman1}
is sufficiently accurate for some alloys \citep{Chen2008}.
However, in ductile metals with second phases, a complex dependence of void nucleation on the
stress state can be observed. For instance,
torsion induces a higher void nucleation
rate than tension in the cast A356 aluminum \citep{Horstemeyer2000}.
In order to account for this effect, the following void nucleation rule was proposed by \cite{HorstemeyerGokhale1999}.
First, recall that $\hat{\mathbf{\Sigma}}$ is the effective stress tensor. Let $J_2$ and $J_3$ be the invariants of
the deviatoric part $\hat{\mathbf{\Sigma}}^{\text{D}}$. The hydrostatic stress component is captured by $\text{tr} \hat{\mathbf{\Sigma}}$.
The void nucleation is governed by
\begin{equation}\label{HorstemeyerGokhale}
\dot{N} = \lambda_{\text{i}} \ N \ \Bigg[ p_1 \Big(\frac{4}{27} - \frac{J^2_3}{J^3_2} \Big) + p_2 \frac{J_3}{J^{3/2}_2}
+ p_3 \frac{ | \text{tr} \hat{\mathbf{\Sigma} |} }{J^{1/2}_2} \Bigg],
\end{equation}
where $p_1$, $p_2$ and $p_3$ are material parameters.

\section*{Appendix B (discussion of void nucleation rules)}

\begin{table}[b]
\caption{Qualitative analysis of the void nucleation rules.}
\begin{tabular}{| l | l | l| l| l|}
\hline
Nucleation rule & Equa-         &  Asymptotic              &  Role of hydro-                   &  Increased nucl. rate      \\
                & tion                 &  behavior       &  static stress                     &     under torsion      \\ \hline \hline
Gurland (1972)                  & \eqref{GurlandEquation}      &   linear growth  &   irrelevant             &   impossible                     \\ \hline
Chu and Needleman (1980)         & \eqref{ChuNeedleman1}       &    saturation &   irrelevant                &   impossible                     \\ \hline
Horstemeyer and   & \eqref{HorstemeyerGokhale}   &   exponential &   sign of hydrostatic                                       &   possible     \\
 Gokhale (1999)                                 &                              &   growth      &   stress is irrelevant        &                \\ \hline
New rule (current study)                    & \eqref{VoidNucleatRule} --   &   linear growth &   sensitive to        &   possible     \\
                              &  -- \eqref{EvolUndCompress}  &          &   hydrostatic stress  &                \\ \hline
\end{tabular}
\label{tab4}
\end{table}

Let us compare the void nucleation rules presented in Appendix A with the new
rule introduced in Section 2.6. This comparison is based on different criteria, such as the
saturation of the void number under monotonic loading, the role played by the hydrostatic stress component, and
the increased nucleation rate under torsion compared to tension.

\emph{Saturation:} It is usually assumed that the voids nucleate at certain inclusions, or, more generally, at second phases
which are sites of local stress raisers. Thus, the total
number of voids is limited by the number of inclusions of a certain type.
Gurland's rule (cf. \eqref{GurlandEquation}) predicts a linear accumulation of voids, without any saturation.
A nearly linear void accumulation is also predicted by the newly constructed relations
\eqref{VoidNucleatRule} -- \eqref{EvolUndCompress} (cf. Figure \ref{fig5}).
If needed, the saturation can be introduced by a simple modification, which is not considered in the current study.
Next, the rule from Chu and Needleman (cf. \eqref{ChuNeedleman1}) is realistic in the sense that the maximum void volume fraction is bounded due
to the saturation of the void nucleation. Finally, the rule from Horstemeyer and Gokhale (cf. \eqref{HorstemeyerGokhale})
predicts even an \emph{accelerated} void nucleation. More precisely, the void number under monotonic
loading increases exponentially. Interestingly, the rule from Horstemeyer and Gokhale says that no voids can be
created in initially void-free material. Indeed, if $N |_{t=0} = 0$ then \eqref{HorstemeyerGokhale} yields $N \equiv 0$.

\emph{The role of hydrostatic stress:}
The influence of the hydrostatic stress component is neglected by both \eqref{GurlandEquation} and \eqref{ChuNeedleman1}.
In contrast to this, the new nucleation rule \eqref{EvolUndTension} is highly sensitive to $\text{tr} \hat{\mathbf{\Sigma}}$
such that for larger hydrostatic stresses increased nucleation rates are predicted.
Interestingly, the rule from Horstemeyer and Gokhale (cf. \eqref{HorstemeyerGokhale})
operates with the absolute value of the
hydrostatic stress component $| \text{tr} \hat{\mathbf{\Sigma}} |$.
Thus, the sign of $\text{tr} \hat{\mathbf{\Sigma}}$ is irrelevant!

\emph{Increased nucleation rate under torsion:}
As already mentioned above, torsion  may induce a higher void nucleation
rate than tension. The classic rules from Gurland as well as Chu and Needleman cannot describe
this effect. By appropriate choice of material parameters, the rule from Horstemeyer and Gokhale and the new rule
introduced in Section 2.6 are capable of reproducing this type of behavior.
Note that different systems of invariants are used in these rules.

The theoretical results are summarized in Table \ref{tab4}.
In conclusion, note that the assumption of linear growth of the void number
is compatible with the experimental data available for the cast A356 aluminium (see Figure \ref{fig5}).

\bibliographystyle{elsarticle-harv}

\end{document}